\documentclass[letter,onecolumn]{IEEEtran}
\usepackage[utf8]{inputenc}
\usepackage{graphicx}
\usepackage{amssymb}
\usepackage[cmex10]{amsmath}
\usepackage{color}
\usepackage{theorem}
\usepackage{pifont}
\usepackage{hyperref}
\usepackage{url}
\usepackage{breakurl}
\usepackage{nicefrac}
\usepackage{booktabs}
\usepackage{bbm}
\usepackage[english]{babel}
\usepackage[english = american]{csquotes}
\usepackage{mathtools}
\usepackage{mathdots}
\MakeOuterQuote{"}
\usepackage[ruled,vlined,titlenumbered]{algorithm2e}
\usepackage{comment}
\usepackage{cite}

\makeatletter
\newcommand{\removelatexerror}{\let\@latex@error\@gobble}
\makeatother

\IEEEoverridecommandlockouts

\newtheorem{theorem}{Theorem}
\newtheorem{definition}{Definition}

\newtheorem{corollary}{Corollary}

\newtheorem{example}{Example}

 \newcommand{\qed}{\hfill \mbox{\raggedright \rule{.07in}{.1in}}}

\newcommand{\Fq}{\ensuremath{\mathbb F_{q}}}

\newcommand{\F}{\ensuremath{\mathbb F}}

\newcommand{\myset}[1]{{#1}}
\newcommand{\intervallincl}[2]{\ensuremath{[#1,#2]}}
\newcommand{\intervallexcl}[2]{\ensuremath{[#1,#2-1]}}

\SetAlgoCaptionSeparator{.} 

\newcommand{\printalgoIEEE}[1]
{\begin{center}
\vspace{1ex}
\scalebox{0.95}{
\removelatexerror
\begin{tabular}{p{0.6\textwidth}}
\begin{algorithm}[H]
 #1
\end{algorithm}
\end{tabular}
}
\vspace{1ex}
\end{center}
}

\DeclareMathOperator{\defi}{def}
\newcommand{\defeq}{\overset{\defi}{=}}

\renewcommand{\bar}{\overline}

\renewcommand{\mod}{\;\textnormal{mod}\;}

\DeclareMathOperator{\RRE}{RRE}

\renewcommand{\vec}[1]{\ensuremath{\mathbf{#1}}}
\newcommand{\Mat}[1]{\ensuremath{\mathbf{#1}}}
\newcommand{\vecelements}[1]{\ensuremath{(#1_0 , #1_1 , \dots , #1_{n-1})}}

\newcommand{\vecelementsArb}[2]{\ensuremath{(#1_0 , #1_1, \dots , #1_{#2-1})}}

\renewcommand{\a}{\vec{a}}
\renewcommand{\b}{\vec{b}}

\newcommand{\y}{\mathbf y}

\newcommand{\A}{\Mat{A}}

\renewcommand{\H}{\mathbf H}

\newcommand{\0}{\vec{0}}

\newcommand{\mycode}[1]{\ensuremath{\mathcal{#1}}}

\newcommand{\codelinearHamming}[1]{\ensuremath{[#1]_q}}
\newcommand{\codelinearHammingqInp}[2]{\ensuremath{[#1]_{#2}}}

\newcommand{\redundECC}[3]{\rho_{#3}\left(#1,#2\right)}
\newcommand{\redundPSMC}[4]{r_{#4}\left(#1,#2,#3\right)}
\newcommand{\SMC}[1]{\ensuremath{#1}-SMC}
\newcommand{\PSMCTwo}[2]{\ensuremath{(#1, #2)}-PSMC}
\newcommand{\PSMCOne}[1]{\ensuremath{#1}-PSMC}
\newcommand{\UnreachTwo}[2]{\ensuremath{(#1, #2)}-UMC}

\newcommand{\blockRRE}[1]{
  \underbrace{1 \; \bullet \;  \dots\;   \bullet}_{#1}
}

\newcommand{\blockOnes}[1]{
  \underbrace{1 \; 1 \;  \dots\;  1}_{#1}
}

\newcommand{\blockCirc}[1]{
  \underbrace{\circ \; \circ \;  \dots\;  \circ}_{#1}
}

\usepackage{pgf}
\usepackage{tikz}
\usetikzlibrary{matrix,chains,positioning,arrows,calc,decorations,plotmarks,patterns, fit,backgrounds}
\usepackage{pgfplots}
\usepgfplotslibrary{external}
\usetikzlibrary{external}                       
\tikzsetexternalprefix{tikzpic/}                   

\pgfplotscreateplotcyclelist{mylist}{red,blue,black,yellow,brown}

\tikzset{
    >=stealth',
    mycircle/.style={circle, draw=black, thick, text width=.1em, minimum height=.8em, text centered, font=\scriptsize},
    mycircle_gray/.style={circle, draw=gray, thick, text width=.1em, minimum height=.8em, text centered, font=\tiny},
    mycircle_small/.style={circle,draw=awgray_dark,fill = awgray_dark, inner sep=0,minimum size=.6em},
    mycircle_small_black/.style={circle,draw=black,fill = black, inner sep=0,minimum size=.6em},
    mybox/.style={rectangle,rounded corners,draw=black, thick,text width=1em,minimum height=4em,minimum width=4em,text centered},
    mybox_small/.style={rectangle,rounded corners,draw=black, thick,text width=1em,minimum height=2em,minimum width=2em,text centered},
    mybox_vec/.style={rectangle,rounded corners,draw=black, thick,text width=1em,minimum height=0.7em, minimum width=4em,text centered},
    mybox_vec_short/.style={rectangle,rounded corners,draw=black, thick,text width=1em,minimum height=0.7em, minimum width=2em,text centered},
    pil/.style={->, thick, shorten <=2pt, shorten >=2pt,},
}

\begin{document}
\title{Codes for Partially Stuck-at Memory Cells}
\author{\IEEEauthorblockN{Antonia Wachter-Zeh and Eitan Yaakobi}\\
\IEEEauthorblockA{Department of Computer Science\\
Technion---Israel Institute of Technology, Haifa, Israel\\
Email: \emph{\{antonia, yaakobi\}@cs.technion.ac.il}
\thanks{Parts of this work have been presented at the \emph{$10$th International ITG Conference on Systems, Communications and Coding (SCC) 2015, Hamburg, Germany} \cite{WachterzehYaakobi_PartiallyStuck_SCC2015}.}
}}
\maketitle

\maketitle

\begin{abstract}
In this work, we study a new model of defect memory cells, called \emph{partially stuck-at memory cells}, which is motivated by the behavior of multi-level cells in non-volatile memories such as flash memories and phase change memories. If a cell can store the $q$ levels $0, 1, \dots, q-1$, we say that it is \emph{partially stuck-at} level~$s$, where $1 \leq s \leq q-1$, if it can only store values which are at least $s$. We follow the common setup where the encoder knows the positions and levels of the partially stuck-at cells whereas the decoder does not.

Our main contribution in the paper is the study of codes for masking $u$ partially stuck-at cells. We first derive lower and upper bounds on the redundancy of such codes. The upper bounds are based on two trivial constructions. We then present three code constructions over an alphabet of size $q$, by first considering the case where the cells are partially stuck-at level $s=1$. The first construction works for $u<q$ and is asymptotically optimal if $u+1$ divides $q$. The second construction uses the reduced row Echelon form of matrices to generate codes for the case $u\geq q$, and the third construction solves the case of arbitrary $u$ by using codes which mask \emph{binary} stuck-at cells. We then show how to generalize all constructions to arbitrary stuck levels. Furthermore, we study the dual defect model in which cells cannot reach higher levels, and show that codes for partially stuck-at cells can be used to mask this type of defects as well. Lastly, we analyze the capacity of the partially stuck-at memory channel and study how far our constructions are from the capacity.
\end{abstract}


\section{Introduction}\label{sec:introduction}
Non-volatile memories such as flash memories and phase change memories (PCMs) have paved their way to be a dominant memory solution for a wide range of applications, from consumer electronics to solid state drives. The rapid increase in the capacity of these memories along with the introduction of multi-level technologies have significantly reduced their cost. However, at the same time, their radically degraded reliability has demanded for advanced signal processing and coding solutions.

The cells of PCMs can take distinct physical states. In the simplest case, a PCM cell has two possible states, an amorphous state and a crystalline state. Multi-level PCM cells can be designed by using partially crystalline states~\cite{B_etal10}. Failures of PCM cells stem from the heating and cooling processes of the cells. These processes may prevent a PCM cell to switch between its states and thus the cells become \emph{stuck-at}~\cite{GPR09,KA05,LJLKC09,PRPOTILB04}. Similarly, in the multi-level setup, a cell can get stuck-at one of the two extreme states, amorphous or crystalline. Or, alternatively, the cells cannot be programmed to a certain state, but can be in all other ones; for example, if a cell cannot be programmed to the amorphous state it can still be at the crystalline state as well as all intermediate ones.
A similar phenomenon appears in flash memories. Here, the information is stored by electrically charging the cells with electrons in order to represent multiple levels. If charge is trapped in a cell, then its level can only be increased, or it may happen that due to defects, the cell can only represent some lower levels. Inspired by these defect models, the goal of this paper is the study of codes which mask cells that are \emph{partially stuck-at}.

In the classical version of stuck-at cells (see e.g.~\cite{KT74}), the memory consists of $n$ binary cells of which $u$ are stuck-at either in the zero or one state.
A cell is said to be \emph{stuck-at level $s \in \{0,1\}$} if the value of the cell cannot be changed. Therefore, only data can be written into the memory which matches the fixed values at the stuck-at cells. A \emph{code for stuck-at cells} maps message vectors on codewords which mask the stuck-at cells, i.e., the values of each codeword at the stuck positions coincide with their stuck level.
The encoder knows the locations and values of the stuck-at cells, while the decoder does not. Hence, the task of the decoder is to reconstruct the message given only the codeword.
The challenge in this defect model is to construct schemes where a large number of messages can be encoded and successfully decoded. If $u$ cells are stuck-at, then it is not possible to encode more than $n-u$ bits, and thus the problem is to design codes with redundancy close to $u$. The same problem is relevant for the non-binary setup of this model, where the cells can be stuck at any level.


The study of codes for memories with stuck-at cells, also known as \emph{memories with defects}, takes a while back to the 1970s. To the best of our knowledge, the problem was first studied in 1974 by Kuznetsov and Tsybakov~\cite{KT74}. Since then, several more papers have appeared, e.g.~\cite{BS77, BordenVinck-OnCodingForStuckAtDefects_1987,Chen_LinearCodesMaskingDefects_1985,Heegard-PartitionedLinearBlockCodesStuckAtDefects_1983,KuznetsovKasamiYamamura-AnErrorCorrectingSchemeDefectiveMemory_1978,K85,LKD78,Tsybakov-ReliableComputationAndReliableStorageOfInformation_1975,T75a,T75b}.
The main goal in all these works was to find constructions of codes which mask a fixed number of stuck-at cells and correct another fixed number of additional random errors. Recently, the connection between memories with stuck-at cells and the failure models of PCM cells has attracted a renewed attention to this prior work. Several more code constructions as well as efficient encoding and decoding algorithms for the earlier constructions were studied; see  e.g.~\cite{FGS05,KimKumar-CodingForMemoryWithStuckAtDefects_2013,LastrasJagmohanFranc-AlgorithmsFormemoriesWithStuckCells_2010, LJF10b, SWSRL10}.

This paper studies codes which mask cells that are \emph{partially stuck-at}. We assume that cells can have one of the $q$ levels $0,1,\ldots,q-1$. Then, it is said that a cell is \emph{partially stuck-at} level $s$, where $1\leq s\leq q-1$, if it can only store values which are greater than or equal to $s$.
In this work, we provide upper and lower bounds on the redundancy and give several code constructions with small redundancy in order to mask partially stuck-at cells.
To facilitate the explanations, our constructions are first given for the case $s=1$, and later generalized to arbitrary levels.

In the first part of the paper, we derive lower and upper bounds on the redundancy of codes that mask partially stuck-at memory cells.
The first lower bound is based on the number of states that each partially stuck-at cell can attain. For the case where all cells are partially stuck-at the same level~$s$, we also derive an improved lower bound.
Further, codes which mask any number of partially stuck-at cells can simply be constructed by using only the levels $\max_i s_i, \dots, q-1$ in each cell.
Alternatively, codes which mask $u$ stuck-at cells (not partially) can be used to mask $u$ partially stuck-at cells. 
Thus, these two schemes provide an upper bound on the minimum redundancy of codes that is necessary to mask $u$ partially stuck-at cells.

In the course of this paper, we provide several code constructions and analyze how far they are from our lower and upper bounds on the redundancy. In particular, for $u<q$, we first show that one redundancy symbol is sufficient to mask all partially stuck-at cells.
Further, an improvement of this construction turns out to be asymptotically optimal in terms of the redundancy.
Our second construction uses the parity-check matrix of $q$-ary error-correcting codes. The third construction uses \emph{binary} codes which mask (usual) stuck-at cells. We also show how the codes we propose in the paper can be used for the dual defect model in which cells cannot reach higher levels. Lastly, we analyze the capacity of the partially stuck-at memory channel and study how far our constructions are from the capacity.

The rest of the paper is organized as follows. In Section~\ref{sec:preliminaries}, we introduce notations and formally define the model of partially stuck-at cells studied in this paper.
Our lower and upper bounds on the redundancy are presented in Section~\ref{sec:bounds}.
In Section~\ref{sec:u<q construction}, we propose codes along with encoding and decoding algorithms for the case $u<q$. In Section~\ref{sec:u=q_construction}, we give a construction based on $q$-ary codes and in Section~\ref{sec:binary construction} we show a more general solution by using codes which mask binary stuck-at cells.
Section~\ref{sec:generalized} generalizes the previous constructions to cells which are partially stuck-at arbitrary (possibly different) levels $s_0, s_1, \dots, s_{u-1}$.
The (dual) problem of codes for unreachable levels is studied in Section~\ref{sec:unreachable} and Section~\ref{sec:capacity} analyzes the capacity of the partially stuck-at channel.
Finally, Section~\ref{sec:conclusion} concludes this paper.

\section{Definitions and Preliminaries}\label{sec:preliminaries}

\subsection{Notations}
In this section, we formally define the models of (partially) stuck-at cells studied in this paper and introduce the notations and tools we will use in the sequel.
In general, for positive integers $a,b$, we denote by $[a]$ the set of integers $\{0,1,\ldots,a-1\}$ and $\intervallexcl{a}{b} = \{a,a+1, \dots, b-1\}$.
Vectors and matrices are denoted by lowercase and uppercase boldface letters, e.g. $\a$ and $\A$, and are indexed starting from~$0$. The all-zero and all-one vectors of length $n$ are denoted by $\0_n$ and~$\mathbf{1}_n$. 
Further, for a prime power $q$, $\Fq$ denotes the finite field of order $q$.

We consider $n$ memory cells with $q$ levels, i.e., they are assumed to represent $q$-ary symbols having values belonging to the set $[q]$.
We can therefore represent the memory cells as a vector in $[q]^n$.

In the classical model of stuck-at cells, a cell is said to be {\emph{stuck-at level $s\in [q]$}} if it can store only the value $s$. In our new model of partially stuck-at cells, a cell is {\emph{partially stuck-at level $s\in [q]$}} if it can store only values which are at least $s$. In the first part of this work, when studying partially stuck-at cells, we only consider $s=1$ and we call such cells {\emph{partially stuck-at-${1}$}}.
Throughout this paper, we also use the notation {\emph{partially stuck-at-${s}$}}, when one or several cells are partially stuck-at level $s$, Lastly, in the most general case, $u$ cells are {\emph{partially stuck-at-$\vec{s}$}}, where $\vec{s} = \vecelementsArb{s}{u}$ is a vector and contains the different stuck levels $s_0,s_1,\dots,s_{u-1}$ of the~$u$ defect cells.


\subsection{Definitions for (Partially) Stuck-at Cells}
We use the notation $(n,M)_q$ to indicate a coding scheme to encode $M$ messages into vectors of length $n$ over the alphabet $[q]$. 
Its redundancy is denoted by $r= n-\log_qM$. A linear code over $[q]$ (in which case $q$ is a power of a prime and $[q]$ corresponds to the finite field $\F_q$ of order $q$), will be denoted by $[n,k]_q$, where $k$ is its dimension and its redundancy is $r=n-k$. Whenever we speak about a linear $\codelinearHamming{n,k,d}$ code, then $d$ refers to the minimum Hamming distance of an $[n,k]_q$ code. We also denote by $\redundECC{n}{d}{q}$ the smallest (known) redundancy of any linear code of length $n$ and minimum Hamming distance $d$ over $\Fq$. For the purpose of the analysis and the simplicity of notations in the paper, we let $\redundECC{n}{d}{q}= \infty$ in case $q$ is not a power of a prime.

Codes for (partially) stuck-at cells are defined as follows.
\begin{definition}[Codes for (Partially) Stuck-at Cells]\label{def:stuck_cells}

We define the following code properties:
\begin{enumerate}
\item An $(n,M)_q$ \textbf{$u$-stuck-at-masking code} (\SMC{u})~$\mycode{C}$  is a coding scheme with encoder $\mathcal{E}$ and decoder~$\mathcal{D}$. The input to the encoder~$\mathcal{E}$ is the set of locations~$\{\phi_0, \phi_1,\dots, \phi_{u-1}\}\subseteq [n]$, the stuck levels $s_0,s_1,\dots, s_{u-1} \in [q]$ of some $u \leq n$ stuck-at cells and a message $m\in [M]$. Its output is a vector $\vec{y}^{(m)}\in [q]^n$ which matches the values of the~$u$ stuck-at cells, i.e.,
$$y_i^{(m)} = s_i,\ \forall i \in \{\phi_0,\phi_1,\dots,\phi_{u-1}\},$$
and its decoded value is $m$, that is $\mathcal{D}(\vec{y}^{(m)}) =m$.\\[-1ex]
\item
An $(n,M)_q$ \textbf{${(u, \vec{s})}$-partially-stuck-at-masking code} (\PSMCTwo{u}{\vec{s}})~$\mycode{C}$  is a coding scheme with encoder $\mathcal{E}$ and decoder $\mathcal{D}$. The input to the encoder~$\mathcal{E}$ is the set of locations $\{\phi_0,\phi_1, \dots, \phi_{u-1}\}\subseteq [n]$, the partially stuck levels $\vec{s} = \vecelementsArb{s}{u}\in \intervallexcl{1}{q}^{u}$ of some $u \leq n$ partially stuck-at cells and a message $m\in [M]$. Its output is a vector $\vec{y}^{(m)}\in [q]^n$ which masks the values of the~$u$ partially stuck-at cells, i.e.,
$$y_i^{(m)} \geq s_i, \ \forall i \in \{\phi_0,\phi_1,\dots,\phi_{u-1}\},$$
and its decoded value is $m$, that is $\mathcal{D}(\vec{y}^{(m)}) =m$. \\[-1ex]
\end{enumerate}
\end{definition}

Notice that in contrast to classical error-correcting codes, (P)SMCs are not just a set of codewords, but also an explicit coding scheme with encoder and decoder.
Further, the partially stuck levels stored in $\vec{s}$ have to be known to the decoder of a PSMC, whereas the decoders of SMCs are independent of the stuck levels.
Clearly, a \PSMCTwo{u}{\vec{s}} is also a \PSMCTwo{u^\prime}{\vec{s}^\prime}, where $u^\prime \leq u$ and $\vec{s}^\prime$ is a subvector of $\vec{s}$ of length $u^\prime$.

Whenever we speak about \PSMCTwo{u}{s}s (i.e., $s$ is a scalar), we refer to a code that masks at most $u$ partially-stuck-at-$s$ cells and when simply we speak about a \PSMCOne{u}, we mean a code which masks at most $u$ partially-stuck-at-$1$ cells.


In the design of (P)SMCs, the goal is therefore to minimize the redundancy $n-\log_qM$ for fixed values $u$ and $\vec{s}$, while providing efficient encoding and decoding algorithms. For positive integers $n,q,u$, where $u\leq n$, and a vector $\vec{s} \in \intervallexcl{1}{q}^u$, we denote by $\redundPSMC{n}{u}{\vec{s}}{q}$ the {\emph{minimum redundancy}} of a \PSMCTwo{u}{\vec{s}} over~$[q]$.
Note that throughout this paper we only consider the problem of masking partially stuck-at cells, without correcting additional (random) errors as e.g. in \cite{Heegard-PartitionedLinearBlockCodesStuckAtDefects_1983} for stuck-at cells.

\subsection{Codes for Stuck-at Cells}
Codes for masking stuck-at cells were studied before and one such construction is stated in the following theorem; see e.g.~\cite{Heegard-PartitionedLinearBlockCodesStuckAtDefects_1983}. We show the proof of this theorem, which includes the encoding and decoding of SMCs, for the completeness of results in this paper and since our encoding and decoding algorithms use similar ideas.
\begin{theorem}[Masking Stuck-At Cells, \cite{Heegard-PartitionedLinearBlockCodesStuckAtDefects_1983}]\label{thm:usual-stuck-at}
Let $\mycode{C}$ be an $\codelinearHamming{n,k,d}$ code with minimum distance $d \geq u+1$, where $q$ is a prime power. Then, there exists a \SMC{u} with redundancy $r=n-k$. 
\end{theorem}
\begin{IEEEproof}
The proof is based on explicitly showing in the following two algorithms how to use the error-correcting code~$\mycode{C}$ to mask stuck-at cells. We assume that $\H$ is a systematic $(n-k)\times n$ parity-check matrix of the code $\mycode{C}$ which is known to both, encoder and decoder.

Algorithm~\ref{algo:encoding_usualstuck} describes the encoding process.
We have to prove that in Step~2, a vector $\vec{z}$ always exists such that the encoded vector $\vec{y}$ masks the $u$ stuck-at cells.
Define $\vec{x}= \vecelementsArb{x}{n} = \vec{z} \cdot \H $.
We have to fulfill the following $u$ equations:
\begin{equation*}
w_{\phi_i} +x_{\phi_i} = s_i, \quad \forall i \in [u].
\end{equation*}
Denote by $\H_u$ the $u$ columns of $\H$ indexed by $\phi_0,\dots,\phi_{u-1}$.
Then, we have to find $\vec{z}$ such that
\begin{equation*}
\vec{z} \cdot \H_u = (s_0-w_{\phi_0} , s_1-w_{\phi_1} , \dots , s_{u-1}-w_{\phi_{u-1}}).
\end{equation*}
This is a heterogeneous linear system of equations with at most $u \leq d-1 \leq n-k$ linearly independent equations
and $n-k$ unknowns (the elements of $\vec{z}$). Therefore, there is at least one solution for the vector $\vec{z}$ such that $\vec{y}$ masks the stuck-at cells.

\printalgoIEEE{
\caption{
\newline
\textsc{EncodingStuck}$\big(\vec{m}; \phi_0,\phi_1,\dots,\phi_{u-1}; s_0,s_1,\dots,s_{u-1} \big)$}
\label{algo:encoding_usualstuck}
\DontPrintSemicolon
\SetAlgoVlined
\SetSideCommentRight
\LinesNumbered
\BlankLine
\SetKwInput{KwIn}{\underline{Input}}
\SetKwInput{KwOut}{\underline{Output}}
\SetKwInput{KwIni}{\underline{Initialize}}
\KwIn{
$\bullet$ message: $\vec{m}=(m_0 , m_1 ,\dots, m_{k-1})\in [q]^{k}$
\newline
$\bullet$ positions of stuck-at cells: $\{\phi_0,\phi_1,\dots,\phi_{u-1}\} \subseteq [n]$
\newline
$\bullet$ levels of stuck-at cells: $s_0,s_1,\dots,s_{u-1} \in [q]$
}
\BlankLine

$\vec{w} = (w_0 , w_1, \dots , w_{n-1}) \leftarrow (\mathbf{0}_{n-k} , m_0 , m_1 , \dots m_{k-1})$\\[1ex]

Find $\vec{z} \in [q]^{n-k}$ such that $\vec{y} \leftarrow \vec{w} + \vec{z} \cdot \H$ masks the stuck-at cells\footnote{We assume here and in the rest of the paper that there is a one-to-one mapping $F:[q]\rightarrow \Fq$ so arithmetic operations like multiplication and addition in $\vec{w} + \vec{z} \cdot \H$ are defined as $F(\vec{w}) + F(\vec{z}) \cdot \H$.}\\[1.2ex]
\KwOut{vector $\vec{y}$ with $y_{\phi_i} =s_i$, $\forall i \in \{\phi_0,\dots,\phi_{u-1}\}$}
}

Algorithm~\ref{algo:decoding_usualstuck} describes the decoding process.
\printalgoIEEE{
\caption{
\textsc{DecodingStuck}$\big(\vec{y})$}
\label{algo:decoding_usualstuck}
\DontPrintSemicolon
\SetAlgoVlined
\SetSideCommentRight
\LinesNumbered
\BlankLine
\SetKwInput{KwIn}{\underline{Input}}
\SetKwInput{KwOut}{\underline{Output}}
\SetKwInput{KwIni}{\underline{Initialize}}
\KwIn{
$\bullet$ stored vector: $\vec{y}=(y_0 , y_1 , \dots, y_{n-1}) \in [q]^n$}
\BlankLine
$\widehat{\vec{z}} \leftarrow (y_0, y_1,\dots , y_{n-k-1})$\\[0.8ex]
$\widehat{\vec{w}} =  (\widehat{w}_{0},\widehat{w}_{1}, \dots , \widehat{w}_{n-1})\leftarrow \vec{y} - \widehat{\vec{z}} \cdot \H$\\[0.8ex]
$\widehat{\vec{m}} \leftarrow (\widehat{w}_{n-k} , \widehat{w}_{n-k+1} ,\dots , \widehat{w}_{n-1})$
\BlankLine
\KwOut{message vector $\widehat{\vec{m}}\in [q]^{k}$}
}
We have to prove that $\widehat{\vec{m}} = \vec{m}$.
Note that $\H$ is a \emph{systematic} parity-check matrix and therefore, the first $n-k$ positions of the output vector $\vec{y}$ of Algorithm~\ref{algo:encoding_usualstuck} (which is at the same time the input of Algorithm~\ref{algo:decoding_usualstuck}) equal $\vec{z}$.
Therefore, in Step~1 in Algorithm~\ref{algo:decoding_usualstuck}, we obtain $\widehat{\vec{z}} = \vec{z}$ and thus, in Step~2, $\widehat{\vec{w}} = \vec{y} - \vec{z}\cdot \H = \vec{w}$ and in Step~3, $\widehat{\vec{m}} = \vec{m}$.
\end{IEEEproof}

It is not completely known whether the scheme of Theorem~\ref{thm:usual-stuck-at} is optimal with respect to its achieved redundancy. However, the redundancy of such a code has to be at least $u$ since there are $u$ stuck-at cells that cannot store information.

Let us illustrate Theorem~\ref{thm:usual-stuck-at} (and Algorithms~\ref{algo:encoding_usualstuck} and \ref{algo:decoding_usualstuck}) with an example.
\begin{example}\label{ex:usual-stuck-at}
Let $u=2$, $q=3$ and $n=5$. Therefore, we need a ternary code of minimum distance at least $u+1 = 3$. The largest such code is a $\codelinearHammingqInp{5,2,3}{3}$ code (see~\cite{Grassl:codetables}) with a systematic parity-check matrix
\begin{equation*}
\H=
\begin{pmatrix}
1 & 0& 0& 1& 0\\
0 & 1& 0& 1& 1\\
0 & 0& 1& 0& 1
\end{pmatrix}.
\end{equation*}
We will show how to construct a $(5,3^2=9)_3$ \SMC{2} which masks any $u' \leq u=2$ stuck-at cells.

For the encoding, we proceed as in Algorithm~\ref{algo:encoding_usualstuck}.
Let $\vec{m} = (m_0, m_1)= (2 , 1)$ be the information we want to store and assume that the cells at positions $\phi_0 = 0$ and $\phi_1 = 4$ are stuck-at $s_0 = 1$ and $s_1 = 2$.
Further, let $\vec{w} = (0 , 0 , 0 , m_0 , m_1) = (0 , 0 ,0 , 2 , 1 )$.
Given $\vec{w}$, $\phi_0$, $\phi_1$ and $s_0$, $s_1$, our goal is to find a vector $\vec{x}\in [3]^5$ such that $\vec{y}\equiv(\vec{w} + \vec{x})\bmod 3$ matches the partially stuck-at cells and yet the information stored in~$\vec{m}$ can be reconstructed from $\vec{y}$.

We use $\vec{z}  = (1 , 0 , 1)$. Then, $\vec{x} =\vec{z} \cdot \H= (1, 0 , 1 , 1 , 1)$ and $\vec{y} = \vec{w} + \vec{x} = (1 , 0 , 1 , 0 , 2)$ which masks the stuck-at cells.

To reconstruct $\vec{w}$, given $\y$, we notice that $(y_0 , y_1 , y_2) = \vec{z} = (1 , 0 , 1)$ and we can simply calculate $\vec{y}- (y_0 , y_1 , y_2)\cdot \H$ to obtain $\vec{w}$ and therefore~$\vec{m}$.

Thus, the required redundancy for masking $u=2$ stuck-at cells with Theorem~\ref{thm:usual-stuck-at} is $r=3$ (the first three symbols).
\end{example}

\section{Bounds on the Redundancy}\label{sec:bounds}
For deriving upper and lower bounds on the minimum redundancy of PSMCs, we assume the most general case of \PSMCTwo{u}{\vec{s}}s, where $\vec{s} = \vecelementsArb{s}{u} \in \intervallexcl{1}{q}^u$.
Hence, when a cell at position $\phi_i$ is partially stuck-at level $s_i$, it can only store the values $\{s_i, s_i+1, \dots, q-1\}$.
\begin{theorem}[Bounds on the Redundancy]\label{thm:bounds_stuckatsi}
For any number of $u \leq n$ partially stuck-at cells and any levels $\vec{s} = \vecelementsArb{s}{u} \in \intervallexcl{1}{q}^u$, the value of the minimum redundancy $\redundPSMC{n}{u}{\vec{s}}{q}$ to mask these cells satisfies
\begin{align*}\label{eq:bounds_generalized_si}
& u-\log_q\left(\prod_{i=0}^{u-1}(q- s_i) \right)\\
&\hspace{1.5ex}\leq \redundPSMC{n}{u}{\vec{s}}{q}\\
&\hspace{3ex}\leq\min\left\{ n \cdot \big(1- \log_q\big(q-\max_i\{s_i\}\big)\big),\ \redundECC{n}{u+1}{q} \right\}.
\end{align*}
\end{theorem}
\begin{IEEEproof}
Let us start with the lower bound.
The $n-u$ cells, which are not partially stuck-at, can each carry a $q$-ary information symbol and the $u$ partially stuck-at cells can still represent $q-s_i$ possible values (all values except for $[s_i]$).
Hence, we can store at most $M\leq q^{n-u}\prod_{i=0}^{u-1}(q-s_i)$ $q$-ary vectors and the code redundancy satisfies
\begin{equation*}
r\geq n-\log_q\left(q^{n-u}\prod_{i=0}^{u-1}(q-s_i)\right) =u-\log_q \left(\prod_{i=0}^{u-1}(q- s_i) \right).
\end{equation*}

Second, we prove the upper bound.
A trivial construction to mask any $u \leq n$  partially stuck-at cells is to use only the values $\max_i \{s_i\},\dots,q-1$ as possible symbols in any cell. Any cell therefore stores $\log_q(q-\max_i \{s_i\})$ $q$-ary information symbols. The achieved redundancy is
\[n-\log_q\big((q-\max_i \{s_i\})^n \big)  =  n\big(1- \log_q(q-\max_i \{s_i\})\big),\]
and therefore $\redundPSMC{n}{u}{\vec{s}}{q}\leq n \big(1- \log_q(q-\max_i \{s_i\})\big)$.

Furthermore, every \SMC{u} can also be used as \PSMCTwo{u}{\vec{s}} since the SMC restricts the values of the stuck-at cells more than the PSMC.
With Theorem~\ref{thm:usual-stuck-at}, the redundancy of an SMC is $\redundECC{n}{u+1}{q}$, which provides another upper bound on the value of $\redundPSMC{n}{u}{\vec{s}}{q}$.
\end{IEEEproof}
Note that $\redundECC{n}{u+1}{q} \geq u$ (by the Singleton bound).

The bounds from Theorem~\ref{thm:bounds_stuckatsi} will serve as a reference for our constructions, as we should ensure to construct codes with smaller redundancy than the upper bound and study how far their redundancy is from the lower bound.

For the special case of partially stuck-at-$1$ cells, we obtain the following bounds on the redundancy $\redundPSMC{n}{u}{\vec{1}}{q}$ of a \PSMCOne{u} for all positive integers $n,q,u$ where $u\leq n$:
\begin{align}
&u\left(1 - \log_q(q-1) \right)\nonumber\\ &\hspace{5ex}\leq\redundPSMC{n}{u}{\vec{1}}{q}\label{eq:bounds_pstuckone}\\
&\hspace{11ex}\leq \min\left\{ n \left(1- \log_q(q-1)\right), \; \redundECC{n}{u+1}{q} \right\}.\nonumber
\end{align}
For partially stuck-at-$s$ cells, we further improve the lower bound on the redundancy in the next theorem.

\begin{theorem}[Improved Lower Bound]\label{thm:improve_lower_bound}
For any $(n,M)_q$ \PSMCTwo{u}{s}, we have $M\leq \big\lfloor\frac{q^n+u(q-s)^n}{u+1}\big\rfloor$, and therefore
$$\redundPSMC{n}{u}{s}{q} \geq \log_q(u+1)-\log_q\big(  1+u(1-s/q)^n\big).$$
\end{theorem}
\begin{IEEEproof}
Assume that there exists an $(n,M)_q$ \PSMCTwo{u}{s}, and let $\mathcal{E}$, $\mathcal{D}$ be its encoder, decoder, respectively. In this case we assume that the encoder's input is a message $m\in[M]$ and a set of indices $U\subseteq[n], |U|\leq u$, of the locations of the cells which are partially stuck-at level $s$. We will show that $M\leq \big\lfloor\frac{q^n+(q-s)^n}{u+1}\big\rfloor$. 
For every $m\in [M]$, let
$$\mathcal{D}^{-1}(m) = \{ \vec{y} \in[q]^n\ | \ \mathcal{D}(\vec{y}) =m\}.$$
For every $m\in [M]$ such that $ \mathcal{D}^{-1}(m)\cap ([q]\setminus [s])^n =\emptyset$, we have that $|\mathcal{D}^{-1}(m)|\geq u+1$. To see that, assume on the contrary that $|\mathcal{D}^{-1}(m)| = u$ (the same proof holds if $|\mathcal{D}^{-1}(m)| < u$), so we can write $\mathcal{D}^{-1}(m)= \{\vec{y}_0,\vec{y}_1,\ldots,\vec{y}_{u-1}\}$, while $\vec{y}_j\notin ([q]\setminus [s])^n$ for $j\in [u]$. For $j\in [u]$, let $i_j\in [n]$ be such that $y_{j,i_j}<s$, and $U=\{i_0,i_1,\ldots,i_{u-1}\}$. Then, $\mathcal{E}(m,U)\neq \vec{y}_j$ for all $j\in [u]$ (where the message $m$ and the positions set $U$ are the input to the encoder $\mathcal{E}$) and thus $|\mathcal{D}^{-1}(m)|\geq u+1$, in contradiction.

Since there are $(q-s)^n$ vectors all with values at least $s$, there are $M-(q-s)^n$ vectors of the described type above where for each such vector there exists a message $m$ where $|\mathcal{D}^{-1}(m)|\geq u+1$.
Therefore, we get that $(q-s)^n + (u+1)(M-(q-s)^n) \leq q^n,$
or
$$M\leq \left\lfloor\frac{q^n+u(q-s)^n}{u+1}\right\rfloor.$$
We also conclude that
\begin{align*}
\redundPSMC{n}{u}{s}{q} & \geq n-\log_q\left( \frac{q^n+u(q-s)^n}{u+1} \right) &\\
& = \log_q(u+1)-\log_q\big(  1+u(1-s/q)^n\big).&
\end{align*}
\end{IEEEproof}
For $u=1$, we obtain from Theorem~\ref{thm:improve_lower_bound}:
$$\redundPSMC{n}{1}{s}{q} \geq \log_q2-\log_q\big(  1+(1-s/q)^n\big).$$
For $\vec{s} = \vecelementsArb{s}{u}$ and $s_{max} = \max_{i\in [u]} \{s_i\}$ and $s_{min} = \min_{i\in [u]} \{s_i\}$, we have
\begin{equation}\label{eq:red_comp_diff_s}
\redundPSMC{n}{u}{s_{max}}{q} \geq \redundPSMC{n}{u}{\vec{s}}{q} \geq \redundPSMC{n}{u}{s_{min}}{q},
\end{equation}
so we can always use $\redundPSMC{n}{u}{1}{q}$ as a lower bound for partially stuck-at-$s_0, s_1, \dots,s_{u-1}$ cells.

Examples~\ref{ex:stuck-at-one-ulessq} and~\ref{ex:improved_constructionI} show cases where the lower bound from Theorem~\ref{thm:improve_lower_bound} is better than the lower bound from~\eqref{eq:bounds_pstuckone}.

\section{A Construction for Partially Stuck-at Level $s=1$ Cells for $u<q$}\label{sec:u<q construction}
\subsection{Code Construction}
In this section, we show a simple construction of \PSMCOne{u}s for masking any $u <q$ partially stuck-at-$1$ cells, which works for any $q$ (not necessarily a prime power).
Let us illustrate the idea of this construction with an example.
\begin{example}\label{ex:stuck-at-one-ulessq}
Let $u=2$, $q=3$ and $n=5$. 
We will show how to construct a $(5,3^4=81)_3$ \PSMCOne{u}.

Let $\vec{m} = (m_0 , m_1, m_2 , m_3)= (2 , 0 , 1 , 0)$ be the information we want to store and assume that the two cells at positions $\phi_0 = 1$ and $\phi_1 = 2$ are partially stuck-at-$1$.
We set $\vec{w} = (0 , m_0 ,m_1 , m_2 , m_3) = (0 , 2 , 0 , 1 , 0)$.
Given~$\vec{w}$,~$\phi_0$ and~$\phi_1$, our goal is to find a vector $\vec{x}$ such that $\vec{y}\equiv(\vec{w} + \vec{x})\bmod 3$ masks the partially stuck-at-$1$ cells and yet the information stored in~$\vec{m}$ can be reconstructed from~$\vec{y}$.

We use $\vec{x}=z \cdot \mathbf{1}_5$, for some $z \in [3]$. Since $w_1 = 2$, we can choose any value from $[3]$ for $z$ except for~$1$, and since $w_2=0$, we can choose any value but $0$. Thus, we choose $z = 2$ and encode the vector $\vec{y}$ to be $\vec{y} = \vec{w} + 2 \cdot \mathbf{1}_5 = (2 , 1 , 2 , 0 , 2)$, which  masks the two partially stuck-at-$1$ cells.

To reconstruct $\vec{w}$, given $\y$, notice that $y_0 = z = 2$ and we can simply calculate $\vec{y}-y_0 \cdot \mathbf{1}_5$ to obtain $\vec{w}$ and therefore~$\vec{m}$.

Thus, the required redundancy for masking $u=2$ partially stuck-at-$1$ cells is $r=1$ (the first symbol).
The lower bound from~\eqref{eq:bounds_pstuckone} is $2\cdot(1-\log_32) = 0.738$, the lower bound from Theorem~\ref{thm:improve_lower_bound} gives $0.787$ (and therefore improves upon the bound from Theorem~\ref{thm:bounds_stuckatsi}) and the upper bound is $\min\{5\cdot(1-\log_32),\ \redundECC{5}{3}{3} = 3\}=1.845$, where the second part was shown in Example~\ref{ex:usual-stuck-at}.
\end{example}

The following theorem shows that it is possible to construct a \PSMCOne{u} with this strategy for any $u < q$.
\begin{theorem}[Construction~I for $u<q$]\label{thm:stuck-at-one-onerow}
If $u<q$ and $u\leq n$, then for all $n$, there exists an $(n,M=q^{n-1})_q$ \PSMCOne{u} with redundancy of one symbol.
\end{theorem}
\begin{IEEEproof}
As in the proof of Theorem~\ref{thm:usual-stuck-at}, we give the explicit encoding and decoding algorithms.

The encoding is shown in Algorithm~\ref{algo:encodingI}.
Let ${\phi_0}, {\phi_1}, \dots, {\phi_{u-1}}$ be the positions of some $u$ partially stuck-at-$1$ cells and let $\vec{m}=(m_0, m_1, \ldots, m_{n-2})\in[q]^{n-1}$ be the message to be written to the memory $\vec{w}$. We first set $\vec{w} = (0 , m_0 , m_1 , \ldots , m_{n-2})$. 
Since $u < q $, there exists at least one value $v \in [q]$ such that $w_{\phi_0}, w_{\phi_1}, \dots, w_{\phi_{u-1}} \neq v$.
Choose $z = q-v\equiv -v \;\bmod q$ and therefore, $ w_{\phi_i} + z  \equiv (w_{\phi_i} -v) \bmod q \neq 0$. Thus, the output vector $\vec{y}$ 
masks all $u$ partially stuck-at-$1$ cells.
\printalgoIEEE{
\caption{
\textsc{Encoding-I}$\big(\vec{m}; \phi_0,\phi_1,\dots,\phi_{u-1}\big)$}
\label{algo:encodingI}
\DontPrintSemicolon
\SetAlgoVlined
\SetSideCommentRight
\LinesNumbered
\BlankLine
\SetKwInput{KwIn}{\underline{Input}}
\SetKwInput{KwOut}{\underline{Output}}
\SetKwInput{KwIni}{\underline{Initialize}}
\KwIn{
$\bullet$ message: $\vec{m}=(m_0 , m_1 , \dots, m_{n-2})\in [q]^{n-1}$
\newline
$\bullet$ positions of partially stuck-at-$1$ cells: $\{\phi_0,\phi_1,\dots,\phi_{u-1}\} \subseteq [n]$}
\BlankLine

$\vec{w} = (w_0 , w_1 , \dots , w_{n-1}) \leftarrow (0 , m_0 , m_1 , \dots m_{n-2})$\\[0.8ex]

Set $v\in [q]$ such that $v \notin \{w_{\phi_0}, w_{\phi_1}, \dots, w_{\phi_{u-1}}\}$\\[0.8ex]

$z \leftarrow q-v$\\[0.8ex]

$\vec{y} \leftarrow (\vec{w} + z \cdot \mathbf{1}_n)\bmod q$\\[1.2ex]
\KwOut{vector $\vec{y}$ with $y_{\phi_i}  \geq 1$, $\forall i \in [u]$}
}

The decoding algorithm is shown in Algorithm~\ref{algo:decodingI} and it simply uses the fact that $y_0$ is used to store the value of $z$.
Thus, $\widehat{\vec{m}} = \vec{m}$.

\printalgoIEEE{
\caption{
\textsc{Decoding-I}$\big(\vec{y})$}
\label{algo:decodingI}
\DontPrintSemicolon
\SetAlgoVlined
\SetSideCommentRight
\LinesNumbered
\BlankLine
\SetKwInput{KwIn}{\underline{Input}}
\SetKwInput{KwOut}{\underline{Output}}
\SetKwInput{KwIni}{\underline{Initialize}}
\KwIn{
$\bullet$ stored vector: $\vec{y}=(y_0 , y_1 , \dots, y_{n-1}) \in [q]^n$}
\BlankLine
$\widehat{z} \leftarrow y_0$\\[0.8ex]
$\widehat{\vec{m}} \leftarrow ((y_1 , y_2 ,\dots , y_{n-1})-\widehat{z} \cdot \mathbf{1}_n)\bmod q$
\BlankLine
\KwOut{message vector $\widehat{\vec{m}}\in [q]^{n-1}$}
}

Lastly, since we need only one symbol to store the value of $z=q-v$ in the first cell, the redundancy is one.
\end{IEEEproof}

This principle works for \emph{any} $n$ since $\mathbf{1}_n$ always exists.
Further, $q$ does not have to be a prime power and we can use this strategy even if $[q]$ does not constitute a finite field.

Theorem~\ref{thm:stuck-at-one-onerow} provides a significant improvement compared to codes for the (stronger) model of usual stuck-at cells, where the required redundancy is $\redundECC{n}{u+1}{q} \geq u$. In order to compare our result to the first term in the upper bound from~\eqref{eq:bounds_pstuckone}, we use the approximation
$\ln(q)-\ln(q-1)\approx 1/q$ for $q$ large enough and thus,
$$1-\log_q(q-1)\approx \frac{1}{q\ln q}.$$
Hence, for $n\geq q\ln q$, the construction from Theorem~\ref{thm:stuck-at-one-onerow} achieves smaller redundancy than the upper bound from~\eqref{eq:bounds_pstuckone}. However, for $u<q$ and $n$ large enough the lower bound on the redundancy from Theorem~\ref{thm:improve_lower_bound} is $\log_q(u+1)<1$. This suggests that codes with better redundancy might exist, as we shall see in the next subsection.

\subsection{Improvement of Construction~I}\label{subsec:improvement_ulessq}
In this subsection, we show how to improve Construction~I from Theorem~\ref{thm:stuck-at-one-onerow} and therefore, how to reduce the required redundancy. Let us first show an example of this modification.

\begin{example}\label{ex:improved_constructionI}
Let $u=2$ and $q=6$. Then, with the same principle as in Algorithm~\ref{algo:encoding_usualstuck} and $z \in [3]$, we can mask any two partially stuck-at-$1$ cells with one redundancy symbol.
The important observation here is that $z$ can always be chosen from a small subset of $[q]$.
Assume, the first cell stores the redundancy, namely $y_0 = z$, and takes therefore only three out of six possible values. We can store additional information in this cell e.g. as follows: $0$ is represented by $0$ or $3$; $1$ is represented by $1$ or $4$; and $2$ is represented by $2$ or $5$. By choosing e.g. between storing $1$ or $4$, we store additional information in the redundancy cell. Thus, the information stored (in terms of $q$-ary symbols) increases by $\log_6(6/3)$ and the required redundancy reduces from one symbol to $1-\log_6(2) \approx 0.613$ symbols while the lower bound from~\eqref{eq:bounds_pstuckone} gives $\approx 0.204$. The lower bound from Theorem~\ref{thm:improve_lower_bound} gives $0.284$ for $n=5$, and $0.457$ for $n=10$. For $n \rightarrow \infty$, the lower bound from Theorem~\ref{thm:improve_lower_bound} reaches $0.613$ which is exactly the redundancy achieved by our construction.
\end{example}

In general, assume that $u<q$, then there is some $v  \in [u+1]$ such that $w_{\phi_0}, w_{\phi_1}, \dots, w_{\phi_{u-1}} \neq v$. Hence, if $u$ is significantly smaller than $q$, then there is more than one value to choose the value of $v$ from and
the redundancy decreases by $\log_q(\lfloor q/(u+1) \rfloor)$ symbols.

Thus, with Theorem~\ref{thm:stuck-at-one-onerow} and the previous considerations, we obtain the following result.
\begin{theorem}[Construction~IB]\label{thm:stuck-at-one-onerow-addinfo}
If~$u <q$ and $u \leq n$, then for all $n$, there exists a \PSMCOne{u} over~$[q]$ of length $n$ and redundancy
\begin{equation*}
r = 1-\log_q\left\lfloor \frac{q}{u+1} \right\rfloor.
\end{equation*}
\end{theorem}
\begin{IEEEproof}
In order to prove the statement, we modify Algorithm~\ref{algo:encodingI} to obtain an encoding procedure for the improved construction from Theorem~\ref{thm:stuck-at-one-onerow-addinfo}.
\printalgoIEEE{
\caption{
\textsc{Encoding-Ib}$\big(\vec{m}; m^\prime; \phi_0,\phi_1,\dots,\phi_{u-1}\big)$}
\label{algo:encoding_impr}
\DontPrintSemicolon
\SetAlgoVlined
\SetSideCommentRight
\LinesNumbered
\BlankLine
\SetKwInput{KwIn}{\underline{Input}}
\SetKwInput{KwOut}{\underline{Output}}
\SetKwInput{KwIni}{\underline{Initialize}}
\KwIn{
$\bullet$ message: $\vec{m}=(m_0, m_1, \dots, m_{n-2})\in [q]^{n-1}$
\newline
$\bullet$ additional message: $m^\prime \in \left[\big\lfloor \frac{q}{u+1}\big\rfloor\right]$
\newline
$\bullet$ positions of stuck-at-$1$ cells: $\{\phi_0,\phi_1,\dots,\phi_{u-1}\} \subseteq [n]$
}
\BlankLine

$\vec{w} = (w_0 , w_1 , \dots , w_{n-1}) \leftarrow (0 , m_0 , m_1 , \dots, m_{n-2})$\\[0.8ex]

Set $v \in [u+1]$ such that $v \notin \left\{w_{\phi_0}\bmod (u+1), \dots, w_{\phi_{u-1}}\bmod (u+1)\right\}$\\[1.2ex]

$z \leftarrow q-v-m^\prime(u+1)$\\[0.8ex]

$\vec{y} \leftarrow (\vec{w} + z \cdot \mathbf{1}_n )\bmod q$\\[0.8ex]


\KwOut{vector $\vec{y}$ with $y_{\phi_i} \geq 1$, $\forall i \in [u]$}
}
The encoding algorithm is a generalization of Algorithm~\ref{algo:encodingI}.
We will prove that the output vector $\vec{y}$ masks the partially stuck-at-$1$ cells.
In Step~2, we can always find a value $v \in [u+1]$ as required since the set $\left\{w_{\phi_0}\bmod (u+1), \dots, w_{\phi_{u-1}}\bmod (u+1)\right\}$
has cardinality~$u$ and there are $u+1$ possible values to choose from.
Note that in Step~3, $z \in [q]$ since $v \in [u+1]$ and $m^\prime (u+1)\in [q-u]$.
In Step~4, we obtain:
\begin{equation*}
y_{\phi_i} = (w_{\phi_i}-v-m^\prime(u+1) )\bmod q.
\end{equation*}
Since $ v \neq w_{\phi_i} \bmod (u+1)$ from Step~2, we conclude that $y_{\phi_i} \neq 0$, for all $i \in [u]$, and the partially stuck-at-$1$ positions are masked.

The decoding algorithm has to guarantee that we can recover $\vec{m}$ and $m^\prime$ and is shown in Algorithm~\ref{algo:decoding_improve}.
Here, we notice that $y_0 = z = q-v-m^\prime(u+1)$ from the encoding step and therefore, in Step~1, $\widehat{z} = z$ and in Step~2, $\widehat{v} = v$.
Solving $y_0 = q-v-m^\prime(u+1)$ for $m^\prime$ shows that in Step~3, the result of the fraction is always an integer and that $\widehat{m^\prime} = m^\prime$.
Since $\widehat{z} = z$, it follows that $\widehat{\vec{m}} = \vec{m}$ in Step~4.

\printalgoIEEE{
\caption{
\textsc{Decoding-Ib}$\big(\vec{y})$}
\label{algo:decoding_improve}
\DontPrintSemicolon
\SetAlgoVlined
\SetSideCommentRight
\LinesNumbered
\BlankLine
\SetKwInput{KwIn}{\underline{Input}}
\SetKwInput{KwOut}{\underline{Output}}
\SetKwInput{KwIni}{\underline{Initialize}}
\KwIn{
$\bullet$ stored vector: $\vec{y}=(y_0 , y_1 , \dots, y_{n-1}) \in [q]^n$}
\BlankLine
$\widehat{z} \leftarrow y_0$\\[0.8ex]
$\widehat{v} \leftarrow (q-\widehat{z}) \mod (u+1)$\\[0.8ex]
$\widehat{m^\prime} \leftarrow \dfrac{q-\widehat{z}-\widehat{v}}{u+1}$\\[0.8ex]
$\widehat{\vec{m}} \leftarrow ((y_1,y_2, \dots, y_{n-1})- \widehat{z}\cdot \mathbf{1}_{n-1})\bmod q$
\BlankLine
\KwOut{message vector $\widehat{\vec{m}}\in [q]^{n-1}$
\newline additional message $\widehat{m^\prime}$}
}
\end{IEEEproof}

Clearly, for $u <q$, the redundancy is $r \leq 1$.
The following lemma shows that this construction is asymptotically optimal.
\begin{theorem}[Optimality of Theorem~\ref{thm:stuck-at-one-onerow-addinfo}]\label{lem:optimality_constrI}
If $(u+1)$ divides $q$, the \PSMCOne{u} from Theorem~\ref{thm:stuck-at-one-onerow-addinfo} is asymptotically optimal in terms of the redundancy.
%
\end{theorem}
\begin{IEEEproof}
If $(u+1)|q$, then the redundancy from Theorem~\ref{thm:stuck-at-one-onerow-addinfo} is $r = \log_q(u+1)$.
The lower bound on the minimum redundancy from Theorem~\ref{thm:improve_lower_bound} is $\redundPSMC{n}{u}{\mathbf{1}}{q} = \log_q(u+1)-\log_q\big(  1+u(1-1/q)^n\big)$. The difference between both is
\begin{equation*}
\Delta_q(n,u) = r - \redundPSMC{n}{u}{\mathbf{1}}{q} = \log_q\big(  1+u(1-1/q)^n\big).
\end{equation*}
For $n \rightarrow \infty$ and since $(1-1/q) < 1$, this value approaches
\begin{equation*}
\log_q\big(1+u(1-1/q)^n\big) \rightarrow \log_q\big(1+0\big) = 0.
\end{equation*}
Thus, the construction is asymptotically optimal.
\end{IEEEproof}

\section{Constructions Using $q$-ary Codes for Partially Stuck-at Level $s=1$ Cells}\label{sec:u=q_construction}

For $u \geq q$, it is not always possible to construct \PSMCOne{u}s with the strategy from Theorem~\ref{thm:stuck-at-one-onerow} (and Algorithms~\ref{algo:encodingI}, \ref{algo:decodingI}).
Therefore, this section provides a construction which works well for $u \geq q$ and can be seen as a generalization of Construction~I (instead of the all-one vector, we use a parity-check matrix).
Let us start with a general statement for any $u$ and any prime power $q$.
\begin{theorem}\label{thm:stuck-at-one-and-RRE}
Let $q$ be a prime power and let a $\kappa \times n$ matrix $\H = \big(H_{i,j}\big)^{i \in [\kappa]}_{j \in [n]}$ over $\Fq$ be given.
If the reduced row Echelon (RRE) form of any $\kappa\times u$ submatrix (denoted by $\H^{(u)}$) has the following form (up to column permutations)\footnote{Notice that $\RRE(\H^{(u)})$ can have less than $\kappa$ rows.}:
\begin{align}\label{eq:RRE_stuckatone}
&\RRE(\H^{(u)}) = \\
&\begin{pmatrix}
\smash[b]{\blockRRE{ \leq q-1}} & \circ \;\circ \; \dots \; \circ  &\dots &\dots&\dots \; \circ\\
& \smash[b]{\blockRRE{ \leq q-1}} & \circ \;\circ \; \dots \; \circ& \dots&\dots\; \circ\\
&& \smash[b]{\blockRRE{ \leq q-1}}&\circ\; \dots& \dots\;\circ\\
&&& \ddots & \vdots\\
&\hspace{-10ex}\smash{\clap{\resizebox{0.35cm}{!}{$0$}}}&&& \smash[b]{\blockRRE{ \leq q-1}}\\
\end{pmatrix},\nonumber\\[-0.5ex]\nonumber
\end{align}
where $\bullet$ has to be a non-zero element from $\Fq$ and $\circ$ is any element from $\Fq$,
then, there exists a \PSMCOne{u} over $\Fq$ of length~$n$ and redundancy $r = \kappa$.
\end{theorem}
\begin{IEEEproof}
Assume w.l.o.g. that the partially stuck-at-$1$ positions are $[u]$. 
As before, if $q$ is a prime {power}, we fix a mapping from the elements of the extension field $\Fq$ to the set of integers $[q]$. 
The encoding and decoding follow similar steps as Algorithm~\ref{algo:encoding_usualstuck}.
From a given message vector $\vec{m} \in [q]^{n-\kappa}$, we first define a vector $\vec{w} = (\mathbf{0}_{\kappa}, \vec{m})$. Then, we search a vector $\vec{z}$ such that $\vec{y} = \vec{w} + \vec{z} \cdot \H$ masks all the partially stuck-at-$1$ cells.

First, assume that $\RRE(\H)$ can mask any $u$ partially stuck-at-$1$ cells and let us show that then also $\H$ can mask any $u$ partially stuck-at-$1$ cells.
By assumption, there is a vector $\vec{z} = (z_0 , z_1 ,\dots , z_{\kappa-1})$ such that $\vec{y} = \vec{w} + \vec{z} \cdot \RRE(\H)$ masks all the partially stuck-at-$1$ cells. Since $\H = \Mat{T} \cdot \RRE(\H)$, for some $\kappa \times \kappa$ full-rank matrix~$\Mat{T}$, the vector $\widetilde{\vec{z}} = \vec{z} \cdot \Mat{T}^{-1}$ masks the same stuck-at cells when multiplied by $\H$ since
$\widetilde{\vec{z}}\cdot  \H = \vec{z} \cdot \Mat{T}^{-1}\H = \vec{z} \cdot \RRE(\H)$.

Second, we prove that $\RRE(\H)$ can mask any $u$ partially stuck-at-$1$ cells.
To simplify notations, assume w.l.o.g. that each "block" of $\RRE(\H^{(u)})$ has length exactly $q-1$, if is shorter it is clear that the principle also works.
Similar to the proof of Theorem~\ref{thm:stuck-at-one-onerow}, there is (at least) one value $z_0 \in \Fq$ such that
\begin{equation}\label{eq:req_proof_thmRRE}
z_0 \cdot H_{0,i} \neq - w_i, \ \forall i \in [q-1],
\end{equation}
since~\eqref{eq:req_proof_thmRRE} consists of (at most) $q-1$ constraints and there are $q$ possible values for $z_0$.
Hence, $(z_0 , z_1 , \dots , z_{\kappa-1}) \cdot \RRE(\H^{(u)})$ will mask the first $q-1$ partially stuck-at-$1$ cells for any $z_1, \dots, z_{\kappa-1}$ since $w_i + z_0\cdot H_{0,i}  \neq 0$, for all $i=0,\dots, q-2$.

Similarly, there is (at least) one value $z_1 \in \Fq$ such that
\begin{equation*}
z_1 \cdot H_{1,i} \neq - (w_i+ z_0\cdot H_{0,i}), \ \forall i \in \intervallincl{q-1}{2q-3},
\end{equation*}
and $(z_0 , z_1 , \dots , z_{\kappa-1}) \cdot \RRE(\H^{(u)})$ will mask the second $q-1$ partially stuck-at-$1$ cells for any $z_2, \dots, z_{\kappa-1}$.
This principle can be continued for each block of $q-1$ cells and clearly, column permutations of $\RRE(\H^{(u)})$ pose no problem on this strategy.

\end{IEEEproof}
In contrast to Theorem~\ref{thm:stuck-at-one-onerow}, it is not clear if a similar statement as Theorem~\ref{thm:stuck-at-one-and-RRE} also holds if $q$ is not a prime power.


\begin{example}
Based on Theorem~\ref{thm:stuck-at-one-and-RRE}, the following $\kappa \times n$ matrix, where $\kappa = \lceil\frac{n}{q-1}\rceil$, can clearly mask up to $u = n$ partially stuck-at-$1$ cells over $\Fq$:
\begin{equation*}
\begin{pmatrix}
\smash[b]{\blockOnes{q-1}} & \\
& \smash[b]{\blockOnes{q-1}} & && \hspace{-6ex}\smash{\clap{\resizebox{0.35cm}{!}{$0$}}} \\
&& \smash[b]{\blockOnes{q-1}}& \\
&&& \ddots\\
&\hspace{-11ex}\smash{\clap{\resizebox{0.35cm}{!}{$0$}}}&&& \smash[b]{\blockOnes{\leq q-1}} \\
\end{pmatrix}.\\[2ex]
\end{equation*}
The redundancy of this code for masking any $u\leq n$ stuck-at-one cells is
$r = \kappa = \lceil\frac{n}{q-1}\rceil$.
Thus, for $q \geq n+1$, this provides codes for masking up to $u = n$ partially stuck-at-$1$ cells with redundancy $r = 1$.
However, for $u=n$ partially stuck-at-$1$ cells, we can always use the trivial construction from Section~\ref{sec:preliminaries}, which outperforms the previous matrix for all~$q>2$.
\end{example}

The following theorem shows a construction of \PSMCOne{u}s based on the parity-check matrix of linear codes. The proof uses Theorem~\ref{thm:stuck-at-one-and-RRE}.

\begin{theorem}[Construction~II]\label{thm:construction_u_equal_q}
Let $u \leq q+d-3$, $u\leq n$, $k <n$, and let $\H$ be a systematic $(n-k)\times n$ parity-check matrix of an $\codelinearHamming{n,k,d}$ code.
Then, there exists a \PSMCOne{u} over~$\Fq$ of length $n$ and redundancy $r=n-k$.
\end{theorem}
\begin{IEEEproof}
Since $\H$ is the parity-check matrix of an $\codelinearHamming{n,k,d}$ code, for $\ell \geq d-1$, any $(n-k) \times \ell $ submatrix of $\H$ has rank at least $d-1$ and contains no all-zero column.

Let us consider the reduced row echelon (RRE) of any $u$ columns of $\H$. 
If $u < d-1$, it is a square $u \times u$ matrix of rank $u$ and a special case of \eqref{eq:RRE_stuckatone} with exactly one element in each "block". 
Else, it has at least $d-1$ rows and $u$ columns of the following form:
\begin{equation}\label{eq:matrix_u_equal_q}
\begin{pmatrix}
1 & \circ & \dots &\dots&\dots& \circ \; \circ \;  \dots\;  \circ\\
&1 & \circ & \dots &\dots& \circ \; \circ \;  \dots\;  \circ\\
&&1 & \circ & \dots & \circ \; \circ \;  \dots\;  \circ\\
&&& \ddots& \ddots& \circ \; \circ \;  \dots\;  \circ\\
&\hspace{-1ex}\smash{\clap{\resizebox{0.35cm}{!}{$0$}}}&&& 1 & \smash[b]{\blockCirc{\leq u-d+1}}  \\
\end{pmatrix},\\[2.3ex]
\end{equation}
where $\circ$ is some element from $\Fq$.
The matrix $\H$ contains no all-zero columns, and thus its RRE contains no all-zero columns.
This holds since $\RRE(\H) = \Mat{T} \cdot \H$ for some full-rank matrix~$\Mat{T}$ and therefore, for some column vector $\b^T$ (which denotes an arbitrary column of $\H$), the column vector $\a^T = \Mat{T} \cdot \b^T$ (the corresponding column of $\RRE(\H)$) is all-zero if and only if $\b = \0$.

Hence, each of the (at most) $u-d+1$ rightmost columns in~\eqref{eq:matrix_u_equal_q} has at least one non-zero element. For any of these $u-d+1$ columns, its lowermost non-zero symbol determines to which of the first $d-1$ columns it will be associated. Hence, in the worst case, there is one such set of columns of size $u-d+2$ (the rightmost $u-d+1$ columns and one of the leftmost $d-1$ columns). Thus, for $u-d+2 \leq q-1$,~\eqref{eq:matrix_u_equal_q} is a special case of~\eqref{eq:RRE_stuckatone}.
According to Theorem~\ref{thm:stuck-at-one-and-RRE}, we can therefore mask any $u$ partially stuck-at-$1$ cells with this matrix and
the redundancy is $r=n-k$.

Algorithm~\ref{algo:encodingII} shows the encoding procedure for this \PSMCOne{u}. For simplicity, a \emph{systematic} $(n-k) \times n$ parity-check matrix $\H$ of the $\codelinearHamming{n,k,d}$ code is used.

\printalgoIEEE{
\caption{
\textsc{Encoding-II}$\big(\vec{m}; \phi_0,\phi_1,\dots,\phi_{u-1}\big)$}
\label{algo:encodingII}
\DontPrintSemicolon
\SetAlgoVlined
\SetSideCommentRight
\LinesNumbered
\BlankLine
\SetKwInput{KwIn}{\underline{Input}}
\SetKwInput{KwOut}{\underline{Output}}
\SetKwInput{KwIni}{\underline{Initialize}}
\KwIn{
$\bullet$ message: $\vec{m}=(m_0, m_1 , \dots, m_{k-1})\in [q]^{k}$
\newline
$\bullet$ positions of partially stuck-at-$1$ cells: $\{\phi_0,\phi_1,\dots,\phi_{u-1}\} \subseteq [n]$
}
\BlankLine

$\vec{w} = (w_0, w_1 , \dots , w_{n-1}) \leftarrow (\mathbf{0}_{n-k} , m_0 , m_1 , \dots m_{k-1})$\\[0.8ex]

Find \vec{z} = \vecelementsArb{z}{n-k} as explained in the proof of Theorem~\ref{thm:stuck-at-one-and-RRE}\\[0.8ex]

$\vec{y} \leftarrow \vec{w} + \vec{z} \cdot \H$\\[1.2ex]
\KwOut{vector $\vec{y}$ with $y_{\phi_i}  \geq 1$, $\forall i \in [u]$}
}

The decoding is shown in Algorithm~\ref{algo:decodingII}. Both algorithms work similarly as Algorithms~\ref{algo:encoding_usualstuck} and \ref{algo:decoding_usualstuck}, but the restriction on the error-correcting code $\codelinearHamming{n,k,d}$ is weaker and therefore our \PSMCOne{u} has a smaller redundancy than the \SMC{u}.
\printalgoIEEE{
\caption{
\textsc{Decoding-II}$\big(\vec{y})$}
\label{algo:decodingII}
\DontPrintSemicolon
\SetAlgoVlined
\SetSideCommentRight
\LinesNumbered
\BlankLine
\SetKwInput{KwIn}{\underline{Input}}
\SetKwInput{KwOut}{\underline{Output}}
\SetKwInput{KwIni}{\underline{Initialize}}
\KwIn{
$\bullet$ stored vector: $\vec{y}=(y_0, y_1 , \dots, y_{n-1}) \in [q]^n$}
\BlankLine
$\widehat{\vec{z}} \leftarrow (y_0 ,y_1, \dots, y_{n-k-1})$\\[0.8ex]
$\widehat{\vec{w}} =  (\widehat{w}_{0} , \widehat{w}_{1} ,\dots , \widehat{w}_{n-1})\leftarrow \vec{y} - \widehat{\vec{z}} \cdot \H$\\[0.8ex]
$\widehat{\vec{m}} \leftarrow (\widehat{w}_{n-k} , \widehat{w}_{n-k+1} ,\dots , \widehat{w}_{n-1})$
\BlankLine
\KwOut{message vector $\widehat{\vec{m}}\in [q]^{k}$}
}
\end{IEEEproof}
Notice that for $q=2$, a PSMC is just an SMC and the construction from Theorem~\ref{thm:construction_u_equal_q} is equivalent to the one from Theorem~\ref{thm:usual-stuck-at}.
Theorem~\ref{thm:construction_u_equal_q} leads to the following corollaries.
\begin{corollary}\label{cor:constructionII}
If $q \geq2$ is a prime power, then for all $u \leq n$, there exists a
\PSMCOne{u}
with redundancy $r=\max\{1,\redundECC{n}{u-q+3}{q}\}$ and therefore $\redundPSMC{n}{u}{\mathbf{1}}{q}\leq\max\{1,\redundECC{n}{u-q+3}{q}\} $.
\end{corollary}
\begin{corollary}[Construction~II for $u=q$]\label{cor:constructionu=q}
If $q \geq 2$ is a prime power, then for all $u=q\leq n$, there exists a
\PSMCOne{q}
with redundancy $r=\redundECC{n}{3}{q}$ and therefore $\redundPSMC{n}{q}{\mathbf{1}}{q}\leq\redundECC{n}{3}{q} $.
\end{corollary}
Theorem~\ref{thm:construction_u_equal_q} shows that in many cases, we can decrease the redundancy by several $q$-ary symbols compared to SMCs. More specifically, according to Theorem~\ref{thm:usual-stuck-at}, in order to construct codes which correct $u=q$ stuck-at cells, one needs to use a linear code with minimum distance~$u+1$, while according to Theorem~\ref{thm:construction_u_equal_q}, for $u=q$ partially stuck-at-$1$ cells, it is enough to use a linear code with minimum distance~$3$.
The next example demonstrates that in many cases, this improvement is quite large.
%

\begin{example}
Let $u=q=5$ and $n=30$. To mask $u$ partially stuck-at-$1$ cells, we use the parity-check matrix of a $\codelinearHammingqInp{30,27,3}{5}$ code, which is the code of largest cardinality over $\F_{5}$ for $n=30$ and $d = 3$ and thus its redundancy is $r = 3$ (see~\cite{Grassl:codetables}).

To compare with SMCs, we need a code with minimum distance $d=6$ (compare Theorem~\ref{thm:usual-stuck-at}). The best known one according to~\cite{Grassl:codetables} has parameters $\codelinearHammingqInp{30,22,6}{5}$ so its redundancy is $r=8$. The trivial construction from the upper bound requires redundancy $4.16$. Therefore, we improve upon both constructions.

However, note that it is still far from the lower bounds from~\eqref{thm:bounds_stuckatsi} (which is $0.69$) and from Theorem~\ref{thm:improve_lower_bound} (which is $1.11$).
\end{example}

Another application of Theorem~\ref{thm:construction_u_equal_q} (and Corollary~\ref{cor:constructionu=q}) uses the parity-check matrix of the ternary Hamming code with parameters $\codelinearHammingqInp{n=\frac{3^r-1}{2}, n-r, 3}{3}$ to obtain a \PSMCOne{3}. The redundancy of this construction is $r= \log_3(2n+1)$.

The following example shows how to use the same Hamming code, but double the code length and thus reduce the redundancy.
\begin{example}\label{ex:diff_stuckatone_stuckusual}
Let $u=q= 3$, and $n=8$. We will show how to encode 6 ternary symbols using the matrix
\begin{equation*}
\H =
\begin{pmatrix}
1 & 1& 0 & 0 & 1 & 1 & 1 & 1\\
0 & 0 & 1 & 1& 1 & 1& 2 & 2
\end{pmatrix}.
\end{equation*}
Any two columns of
$\left(\begin{smallmatrix}
 1& 0 &   1 & 1 \\
 0 & 1 &  1& 2
\end{smallmatrix}\right)$
are linearly independent (it is the parity-check matrix of the \codelinearHammingqInp{4,2,3}{3} Hamming code), and therefore, any submatrix of $\H$, which consists of three columns, has rank two and its RRE is of the form stated in~\eqref{eq:RRE_stuckatone}.

Let $\vec{m} = (m_0, m_1, m_2, m_3, m_4, m_5) = (1, 0, 2, 0, 1, 2)$ be the message and assume that the cells at positions $0,2,4$ are partially stuck-at-$1$. We first set the memory state to be $\vec{w} = (0, 0, m_0, m_1, m_2, m_3, m_4, m_5) = (0, 0, 1, 0, 2, 0, 1, 2)$. Similar to the construction in Theorem~\ref{thm:stuck-at-one-onerow}, we seek to find a length-2 ternary vector $\vec{z}=(z_0 , z_1)$ such that the vector $\vec{y} = \vec{w} + \vec{z} \cdot \H$ masks the three partially stuck-at-$1$ cells. Hence, we can choose $\vec{z} = (1 , 1)$ and thus, $\vec{y} = (1, 1, 2, 1, 1, 2, 1, 2)$ masks the three partially stuck-at-$1$ cells. It is possible to show that this property holds for any three partially stuck-at-$1$ cells and any message vector of length~6.

This provides a code of length $n=8$ and redundancy $r = 2$ for masking any $u=3$ partially stuck-at-$1$ cells. The first part of the upper bound from~\eqref{eq:bounds_pstuckone} uses the trivial construction and gives $2.95$. The second part of the upper bound is $\redundECC{8}{4}{4}=4$ (compare~\cite{Grassl:codetables}) and thus, our construction requires less redundancy. The lower bound on the redundancy from~\eqref{eq:bounds_pstuckone} is $1.107$, and the one from Theorem~\ref{thm:improve_lower_bound} is $1.161$.



\end{example}

Example~\ref{ex:diff_stuckatone_stuckusual} is not a special case of Theorem~\ref{thm:construction_u_equal_q}.
The principle from Example~\ref{ex:diff_stuckatone_stuckusual} works for $u=3$ and arbitrary $q$ with the corresponding $q$-ary Hamming code.
Unfortunately, it is not clear how to generalize it to arbitrary values of $u$ or how to find other non-trivial matrices which fulfill the requirements of Theorem~\ref{thm:stuck-at-one-and-RRE}.


\section{Construction Using Binary Codes for Partially Stuck-at Level $s=1$ Cells}\label{sec:binary construction}
In this section, we describe a construction
of \PSMCOne{u}s
for masking~$u$ partially stuck-at-$1$ cells by means of \emph{binary} SMCs. This construction works for any $u$; however, for $u< q$, the constructions from the previous sections achieve smaller redundancy, therefore and unless stated otherwise, we assume in this section that $n \geq u\geq q$.

For a vector $\vec{w}\in [q]^n$ and a set $\myset{U}\subseteq [n]$, the notation $\vec{w}_{\myset{U}}$ denotes the subvector of $\vec{w}$ of length $|U|$ which consists of the positions in~$\myset{U}$. Let $\myset{U}$ contain the locations of the $u$ partially stuck-at-1 cells, and let a vector $\vec{w}\in [q]^n$ be given, then our construction of PSMCs has two main parts:
\begin{enumerate}
\item Find $z \in [q]$ such that the number of zeros and $(q-1)$s is minimized in the vector $$(\vec{w}^{(z)})_{\myset{U}} = (\vec{w} + z \cdot \mathbf{1}_{n+1})_{\myset{U}},$$ and denote this value by $\widetilde{u}$.
\item Use a binary \SMC{\widetilde{u}} to mask these $\widetilde{u}$ stuck-at cells.
\end{enumerate}
This idea provides the following theorem.

\begin{theorem}[Construction~III Using Binary Codes]\label{thm:constr_binary_code}
Let $n$, $q \geq 4$ and $u \leq n $ be positive integers and let $\widetilde{u} = \lfloor {2u}/{q}\rfloor $. Assume that $\widetilde{\mycode{C}}$ is an $\codelinearHammingqInp{n,k,d\geq \widetilde{u}+1 }{2}$ binary \SMC{\widetilde{u}} with encoder $\widetilde{\mycode{E}}$ and decoder $\widetilde{\mycode{D}}$ given by Theorem~\ref{thm:usual-stuck-at}.
Then, there exists a \PSMCOne{u} over $[q]$ of length $(n+1)$ and redundancy
$$r =  (n-k-1)\log_q\left(\frac{q}{\lfloor q/2\rfloor}\right) + 2.$$
\end{theorem}
\begin{IEEEproof}
We will give the explicit algorithms for encoding and decoding. We assume that $\H$ is a systematic $(n-k) \times n$ parity-check matrix of the code $\widetilde{\mycode{C}}$ and we refer to Algorithms~\ref{algo:encoding_usualstuck} and \ref{algo:decoding_usualstuck} as the encoder $\widetilde{\mycode{E}}$ and decoder $\widetilde{\mycode{D}}$ of the code $\widetilde{\mycode{C}}$, respectively.

Let us start with the encoding algorithm for the \PSMCOne{u}.
\printalgoIEEE{
\caption{
\textsc{Encoding-III}$\big(\vec{m}; \vec{m}^\prime;U\big)$}
\label{algo:encoding_binary}
\DontPrintSemicolon
\SetAlgoVlined
\SetSideCommentRight
\LinesNumbered
\BlankLine
\SetKwInput{KwIn}{\underline{Input}}
\SetKwInput{KwOut}{\underline{Output}}
\SetKwInput{KwIni}{\underline{Initialize}}
\KwIn{
$\bullet$ messages: $\vec{m}=(m_0, m_1 , \dots , m_{k-1})\in [q]^{k}$ and \newline \phantom{$\bullet$ }$\vec{m'}=(m_0' , m_1' , \dots , m_{n-k-2}')\in \big[\lfloor q/2\rfloor\big]^{n-k-1}$
\newline
$\bullet$ positions of partially stuck-at-$1$ cells:\\ \hspace{8ex} $U = \{\phi_0,\phi_1,\dots,\phi_{u-1}\} \subseteq [n+1]$}
\BlankLine

$\vec{w} = (w_0 , w_1 , \dots , w_{n}) \leftarrow (\mathbf{0}_{n-k} , m_0 , m_1 , \dots , m_{k-1} , 0)$\\[0.8ex]

Find $z\in [q]$ such that the vector $(\vec{w}^{(z)})_U$ has at most $\widetilde{u}$ zeros and $(q-1)$s, where $\vec{w}^{(z)} = (\vec{w} + z\cdot \vec{1}_{n+1}) \bmod q$. Let $\widetilde{U}$ be the set of these positions. \\[0.8ex]

If $z>0$ then $w_n^{(z)} \leftarrow z$, else $w_n^{(z)} \leftarrow q-2$\\[0.8ex]

Let $\vec{v} = (v_0 , v_1 , \dots , v_{n-1}) \in [2]^n$ be such that $v_i \leftarrow 1$ if $w^{(z)}_i = q-1$, else $v_i\leftarrow 0$ \\[0.8ex]

Apply Algorithm~\ref{algo:encoding_usualstuck}: $\vec{c'} \leftarrow\widetilde{\mycode{E}}\big((v_{n-k},v_{n-k+1}, \ldots , v_{n-1});\widetilde{U};\mathbf{1}_n\big)$\newline
$\widetilde{\vec{c}} \leftarrow (\vec{c'} \ 0)$\\[0.8ex]

$\widetilde{\vec{m}} \leftarrow (2m_0' , 2m_1' , \dots , 2m_{n-k-2}' ,\mathbf{0}_{k+2} ) \in [q]^{n+1}$\\[0.8ex]


$\vec{y} \leftarrow (\vec{w}^{(z)}  + \widetilde{\vec{c}} + \widetilde{\vec{m}}) \bmod q$\\[1.2ex]
\BlankLine
\KwOut{vector $\vec{y}\in [q]^{n+1}$ with $y_{\phi_i} \geq 1$, $\forall i \in [u]$}
\BlankLine
}


First, we show that in Step~2 of Algorithm~\ref{algo:encoding_binary}, there exists $z\in[q]$ such that the vector $(\vec{w}^{(z)})_U$ has at most $\widetilde{u}$ zeros and $(q-1)$s, where $\vec{w}^{(z)} = (\vec{w} + z\cdot \vec{1})\bmod q$. The value $\widetilde{u}$ is equivalent to the minimum number of cells in any two (cyclically) consecutive levels in $\vec{w}$. That is, let us denote  $v_i = |\{j: w_j = i, j \in \{\phi_0, \dots, \phi_{u-1}\}|$, for all $i \in [q]$. Then, we seek to minimize the value of
$$v_i+v_{(i+1)\bmod q},$$ where $i \in [q]$.
Since $\sum_{i=0}^{q-1} (v_i +v_{(i+1)\bmod q}) = 2u$, according to the pigeonhole principle, there exists an integer $z\in [q]$ such that $v_z +v_{(z+1)\bmod q} \leq \lfloor {2u}/{q}\rfloor = \widetilde{u}$. Therefore, the vector $(\vec{w}^{(z)})_{U}$ has at most $\widetilde{u}$ zeros and $(q-1)$s, and the set of these positions is denoted by ~$\widetilde{U}$. We write the value of $z$ in the last cell and if $z=0$, we write $w_n=q-1$ which is necessary in the event where the last cell is partially stuck-at-1.

Next, we treat the $\widetilde{u}$ cells of $\vec{w}^{(z)}$ of value $0$ or $q-1$ as binary stuck-at cells since adding $0$ or $1$ to the other $u-\widetilde{u}$ cells still masks those partially stuck-at-$1$ cells as they will not reach level 0. Therefore, according to Theorem~\ref{thm:usual-stuck-at}, we can use the encoder of $\widetilde{\mycode{E}}$ of the code $\widetilde{\mycode{C}}$ as done in Steps 4 and 5. The first part of Step~5 is a call of Algorithm~\ref{algo:encoding_usualstuck}.
In order to do so, we first generate a length-$n$ binary vector $\vec{v}$ where its value is~$1$ if and only if the corresponding cell has value $q-1$. Then, we require that the $\widetilde{u}$ cells will be stuck-at level 1. This will guarantee that the output $\widetilde{\vec{c}}$ has value 1 if the corresponding partially stuck-at cell was in level 0 and its value is 0 if the corresponding cell was in level $q-1$. We note that here we took only the last $n-k$ indices of the vector $\vec{v}$ as this is the form of the input vector in Algorithm~\ref{algo:encoding_usualstuck}, while the vector $\vec{v}$ corresponds to the vector $\vec{w}$ in Algorithm~\ref{algo:encoding_usualstuck}.

The first $n-k$ cells contain so far only binary values, thus, we use the first $n-k-1$ cells to write another symbol with~$\lfloor q/2\rfloor$ values in each cell, as done in Steps 6 and 7. The last cell in this group of $n-k$ cells remains to store only a binary value, which is necessary to determine the value of $z$ in the decoding algorithm.

To complete the proof, we present the decoding algorithm.
\printalgoIEEE{
\caption{
\textsc{Decoding-III}$\big(\vec{y})$}
\label{algo:decoding_binary}
\DontPrintSemicolon
\SetAlgoVlined
\SetSideCommentRight
\LinesNumbered
\BlankLine
\SetKwInput{KwIn}{\underline{Input}}
\SetKwInput{KwOut}{\underline{Output}}
\SetKwInput{KwIni}{\underline{Initialize}}
\KwIn{
$\bullet$ stored vector: $\vec{y}=(y_0 , y_1 , \dots, y_{n}) \in [q]^{n+1}$}
\BlankLine
If $(y_{n-k-1}-y_n)\bmod q \leq 1$ then $\widehat{z} \leftarrow y_n$, else $\widehat{z} \leftarrow 0$\\[0.8ex]
$\widehat{\vec{y}} \leftarrow \vec{y} - \widehat{z} \cdot \mathbf{1}_{n+1}$\\[0.8ex]
$\widehat{\vec{m'}}\leftarrow (\left\lfloor\widehat{y}_0/2 \right\rfloor,\left\lfloor\widehat{y}_0/2 \right\rfloor, \dots, \left\lfloor\widehat{y}_{n-k-2}/2\right\rfloor)$\\[0.8ex]
$\widehat{\vec{t}}\!\leftarrow\! (\widehat{y}_0-2 \widehat m'_0, \dots , \widehat{y}_{n-k-2}-2 \widehat m'_{n-k-2} , \widehat{y}_{n-k-1})\bmod q$\\[0.8ex]
$\widehat{\vec{c'}} \leftarrow \widehat{\vec{t}} \cdot \H$\\[0.8ex]
$\widehat{\vec{m}} \leftarrow (\widehat{y}_{n-k} -\widehat c'_{n-k} , \dots, \widehat{y}_{n-1} - \widehat c'_{n-1})\bmod q$
\BlankLine
\KwOut{message vectors $\widehat{\vec{m}}\in [q]^{k}$ and $\widehat{\vec{m'}} \in \big[\lfloor q/2\rfloor\big]^{n-k-1}$}
\BlankLine
}

In Step 1, we first determine the value of $z$. Note that if $z\neq 0$ then $y_{n-k}= y_n =z$ or $y_{n-k}= (y_n+1)\bmod q$ and in either case the condition in this step holds. In case $z=0$ then $y_n=q-2$ and $y_{n-k}=0$ or $y_{n-k}=1$ so $(y_{n-k-1}-y_n)\bmod q \geq 2$ (since $q \geq 4$). Therefore, we conclude that $\widehat{z}=z$.

In Step 3, we determine the value of the second message $\vec{m'}$. Since $\widehat{z}=z$, we get that $\widehat{\vec{m'}} = \vec{m'}$. In Steps 4 and 5, we follow Algorithm~\ref{algo:decoding_usualstuck} in order to decode the vector $\vec{c'}$ and get that $\widehat{\vec{c'}} = \vec{c'}$. Lastly, in Step 5 we retrieve the first message vector and get $\widehat{\vec{m}} = \vec{m}$.

The size of the message we store in this code is $M=q^k \cdot \lfloor (q/2)\rfloor^{n-k-1}$ and the redundancy of this \PSMCOne{u} is
\begin{align*}
 r & = n  + 1 - \log_qM = n+1- \log_q (q^k \cdot \lfloor q/2\rfloor^{n-k-1})\\
& = n-k +1 - (n-k-1)\cdot \log_q  (\lfloor q/2\rfloor) \\
& = (n-k-1) \log_q \left(\frac{q}{\lfloor q/2\rfloor}\right) +2.
\end{align*}
\end{IEEEproof}

The following examples shows the complete encoding and decoding processes described in Theorem~\ref{thm:constr_binary_code}.

\begin{example}
Let $q=4$, $n=15$, $k=11$, $u=5$ and $U = \{1,4,8,12,15\} \subset [16]$ be the set of partially stuck-at-$1$ positions.
Then, $\widetilde{u} = \left \lfloor 2u/q\right \rfloor = 2$ and we can use the \codelinearHammingqInp{15, 11, 3}{2} code $\widetilde{\mycode{C}}$ as the binary \SMC{2}. The following matrix is a systematic parity-check matrix of $\widetilde{\mycode{C}}$:
\begin{equation*}
\H =
\left(
\begin{smallmatrix}
1 &0 &0 &0 &0 &0 &0 &0 &0 &1 &1 &1 &1 &1 &1\\
0 &1 &0 &0 &0 &0 &1 &1 &1 &0 &0 &0 &1 &1 &1\\
0 &0 &1 &0 &1 &1 &0 &1 &1 &0 &1 &1 &0 &0 &1\\
0 &0 &0 &1 &0 &1 &1 &0 &1 &1 &0 &1 &0 &1 &1\\
\end{smallmatrix}\right).
\end{equation*}
Let $\vec{m}  = (0 , 3 , 2 , 1 , 2 , 2 , 3 , 1 , 3 , 2 , 2 )$ and $\vec{m}^\prime = (1 , 0 , 1)$.

Let us first show the steps of the encoding algorithm.
With $z=1$ in Step~2, we have in the different steps of Algorithm~\ref{algo:encoding_binary} (the partially stuck-at-$1$ positions are underlined):
\begin{align*}
&\text{Step~1: }\qquad\vec{w} = (0 , \underline{0} , 0 , 0 , \underline{0} , 3 , 2 , 1 , \underline{2} , 2 , 3 , 1 , \underline{3} , 2 , 2 , \underline{0})\\
&\text{Steps~2\&3: }\;\vec{w}^{(z)} = (1 , 1 , 1 , 1 , 1 , 0 , 3 , 2 , 3 , 3 , 0 , 2 , 0 , 3 , 3 , 1)
\end{align*}
Therefore, $(\vec{w}^{(z)})_U = (1 , 1 , 3 , 0 , 1)$ has two $0$s/$3$s which equals $\widetilde{u}$. Further, we obtain:
\begin{align*}
\text{Step~4: }\qquad\,\vec{v}& = (0 , 0 , 0 , 0 , 0 , 0 , 1 , 0 , 1 , 1 , 0 , 0 , 0 , 1 , 1 )\\
\text{Step~5: }\qquad\vec{c}^\prime & = (1 , 0 , 0, 0) \cdot \H=\\
& = (1 , 0 , 0 , 0 , 0 , 0 , 0 , 0 , 0 , 1 , 1 , 1 , 1 , 1 , 1)\\
\widetilde{\vec{c}}& = (1 , 0 , 0 , 0 , 0 , 0 , 0 , 0 , 0 , 1 , 1 , 1 , 1 , 1 , 1 , 0)\\
\text{Step~6: }\!\qquad\widetilde{\vec{m}}& = (2 , 0 , 2 , 0 , 0 , 0 , 0 , 0 , 0 , 0 , 0 , 0 , 0 , 0 , 0 , 0)\\
\text{Step~7: }\,\qquad{\vec{y}}& = (0 , \underline{1} , 3 , 1 , \underline{1} , 0 , 3 , 2 , \underline{3} , 0 , 1 , 3 , \underline{1} , 0 , 0 , \underline{1})
\end{align*}
Clearly, the partially stuck-at-$1$ positions are masked in the vector $\vec{y}$.

Let us now show the decoding process, i.e., Algorithm~\ref{algo:decoding_binary}.
\begin{align*}
\text{Step~1: } y_3-&y_{15} = 0 \Longrightarrow \widehat{z} = 1\\
\text{Step~2: }\hspace{2.5ex} \widehat{\vec{y}}& = (3 , 0 , 2 , 0 , 0 , 3 , 2 , 1 , 2 , 3 , 0 , 2 , 0 , 3 , 3 , 0)\\
\text{Step~3: }\hspace{1ex}\widehat{\vec{m}^\prime}& = (\left\lfloor 3/2 \right\rfloor ,0 , \left\lfloor 2/2\right\rfloor) = (1 , 0 , 1)\\
\text{Step~4: }\hspace{3ex}\widehat{\vec{t}}& = (3-2 , 0-0 , 2-2 , 0) = (1 , 0 , 0 , 0)\\
\text{Step~5: }\hspace{2.2ex}\widehat{\vec{c}^\prime}& =\widehat{\vec{t}} \cdot \H =  (1 , 0 , 0 , 0 , 0 , 0 , 0 , 0 , 0 , 1 , 1 , 1 , 1 , 1 , 1)\\
\text{Step~6: }\hspace{1.7ex}\widehat{\vec{m}}& = (0 , 3 , 2 , 1 , 2 , 2 , 3 , 1 , 3 , 2 , 2 ).
\end{align*}
We have therefore successfully recovered the messages $\vec{m}$ and~$\vec{m}^\prime$ and
the redundancy to mask these five partially stuck-at-$1$ cells is $r = 3.5$ $q$-ary cells. Construction~II from Corollary~\ref{cor:constructionII} requires redundancy $\redundECC{16}{4}{4} = 4$.

The lower bound from~\eqref{eq:bounds_pstuckone} gives $1.037$ and the lower bound from Theorem~\ref{thm:improve_lower_bound} gives $1.26$.

As a comparison, to mask $u=5$ usual stuck-at cells in a block of $16$ cells, we need a {quaternary} code of length $n=16$ and distance $d \geq u+1 = 6$. The largest such code is a $\codelinearHammingqInp{16,9,6}{4}$ code with 7 redundancy symbols.
The trivial construction from Theorem~\ref{thm:bounds_stuckatsi} needs redundancy $3.11$.
For this example, the new construction is thus worse than the trivial one. However, for larger $n$, the influence of the $+2$ in the redundancy diminishes and our construction outperforms the trivial one (see Example~\ref{ex:binary_constr_2}), but the principle of the construction is easier to show with short vectors.
\end{example}

\begin{example}\label{ex:binary_constr_2}
Let $q=4$, $n=63$, $k=57$ and $u=5$.
Then, the required redundancy for
Construction~II from Corollary~\ref{cor:constructionII} is $\redundECC{64}{4}{4} = 6$ and for Construction~III from Theorem~\ref{thm:constr_binary_code}, it is $r=4.5$. This improves significantly upon the upper bounds from Theorem~\ref{thm:bounds_stuckatsi} since the trivial construction from Theorem~\ref{thm:bounds_stuckatsi} needs redundancy $13.1$ and $\redundECC{64}{6}{4} = 13$.
\end{example}

The next corollary summarizes this upper bound on the redundancy for $n$ cells.
\begin{corollary}
For all $4 \leq q\leq u\leq n$ we have that
$$\redundPSMC{n}{u}{\vec{1}}{q} \leq \left(\redundECC{n-1}{\big\lfloor \tfrac{2u}{q}\big\rfloor+1}{2}\hspace{-0.5ex}-\hspace{-0.5ex}1\right)\cdot\log_q\left(\frac{q}{\lfloor q/2\rfloor}\right)+2.$$
\end{corollary}

Therefore, if we use a binary $\codelinearHammingqInp{n=2^{r_H}-1, n-r_H, 3}{2}$ Hamming code as a \SMC{\widetilde{u}} with $\widetilde{u} = 2$, then $u \leq q+\left\lfloor\frac{q-1}{2}\right\rfloor$ has to hold. For even $q$, we therefore have $u\leq q+q/2-1$. Then, the required redundancy is $r = (r_H-1)\log_q(2) +2= (\log_2(n+1)-1)\cdot\log_q(2)+2$.

Note that for $u \rightarrow n$ and for large $q$, the trivial construction from Section~\ref{sec:preliminaries} is quite good. Construction~III outperforms the trivial construction if $u > q$ and $n \gg u$.

\section{Generalization of the Constructions to Arbitrary Partially Stuck-at Levels}\label{sec:generalized}

The main goal of this section is to consider the generalized model of partially stuck-at cells as in Definition~\ref{def:stuck_cells}. This model is applicable in particular for non-volatile memories where the different cells might be partially stuck-at different levels.

Let $U= \{\phi_0,\phi_1,\dots,\phi_{u-1}\}$ be the set of locations of the partially stuck-at cells where the $i$-th cell, for $i \in \{\phi_0,\phi_1,\dots,\phi_{u-1}\}$, is partially stuck-at level $s_i$. We denote by $u_i$, $i\in [q]$, the number of cells which are partially stuck-at level $i$ (thus, $\sum_{i=1}^{q-1} u_i = u$) and by $\myset{U}_{i}\subseteq [n]$, $\forall i \in [q]$, the set of positions which are partially stuck-at-$i$. Note that some values of $u_i$ might be zero and thus the corresponding sets $\myset{U}_i$ are the empty set.

In the sequel of this section, we generalize the constructions from Sections~\ref{sec:u<q construction}, \ref{sec:u=q_construction} and \ref{sec:binary construction} to the generalized model of partially stuck-at-$\vec{s}$ cells, where $\vec{s} = \vecelementsArb{s}{u} \in \intervallexcl{1}{q}^u$.
We only give the main theorems without showing the explicit decoding algorithms and mostly without examples since these results are a direct consequence of the previous sections.

\subsection{Generalization of Construction~I}
Construction~I from Theorem~\ref{thm:stuck-at-one-onerow} (including its improvement from Theorem~\ref{thm:stuck-at-one-onerow-addinfo}) generalizes as follows.

\begin{theorem}[Generalized Construction~I]\label{thm:stuck-at-one-onerow_Di}
If $\sum_{i=0}^{u-1}s_i < q$ and $u \leq n$, then for all $n$,
there exists a \PSMCTwo{u}{\vec{s}}, where $\vec{s} = \vecelementsArb{s}{u} \in \intervallexcl{0}{q}^u$, over $[q]$ of length $n$ and redundancy
\begin{equation*}
r = 1-\log_q\left\lfloor\frac{q}{\sum_{i=0}^{u-1}s_i+1}\right\rfloor.
\end{equation*}
\end{theorem}
\begin{IEEEproof}
The proof is a generalization of the proofs of Theorem~\ref{thm:stuck-at-one-onerow} and Theorem~\ref{thm:stuck-at-one-onerow-addinfo}.
Let $\vec{m} \in [q]^{n-1}$ be the message vector, define $\vec{w} = (0 \ \vec{m})$ and let $\vec{y} = \vec{w} + z \cdot \mathbf{1}_n$ be the vector that we will store.
We have to find $z$ such that $y_{\phi_i}  = (w_{\phi_i}+z) \notin [s_i] $, $\forall i\in [u]$. Each partially stuck-at-$s_i$ cell excludes at most $s_i$ possible values for $z$, in total the $u$ cells therefore exclude at most $\sum_{i=0}^{u-1}s_i$ values and if $\sum_{i=0}^{u-1}s_i<q$, there is at least one value $z \in [q]$ such that $y_{\Phi_i}  = (w_{\Phi_i}+z) \notin [s_i] $, $\forall i\in [u]$.

In particular, there is always a $z  \in [\sum_{i=0}^{u-1}s_i+1]$ 
such that $y_{\phi_i} = (w_{\phi_i} + z) \notin [s_i]$, for all $i \in [u]$.
Therefore, we can save additional information in the redundancy cell and as a generalization of Theorem~\ref{thm:stuck-at-one-onerow-addinfo}, the statement follows.
\end{IEEEproof}

The encoding and decoding processes are analogous to Algorithms~\ref{algo:encodingI} and~\ref{algo:decodingI}, and Algorithms~\ref{algo:encoding_impr} and \ref{algo:decoding_improve}, respectively. Unfortunately, we cannot claim (as in Theorem~\ref{lem:optimality_constrI} for $s_i=1$, $\forall i$) that this construction is asymptotically optimal, since the lower bound on the redundancy in Theorem~\ref{thm:improve_lower_bound} does not depend asymptotically on the value of $s$. For example, if $s_i=s$, $\forall i$, and $(su +1) \mid q$, then the redundancy of Theorem~\ref{thm:stuck-at-one-onerow_Di} is $r = \log_q(su+1)$, but the lower bound form Theorem~\ref{thm:improve_lower_bound} approaches only $\redundPSMC{n}{u}{s}{q} = \log_q(u+1)$ for $n$ large enough.

\subsection{Generalization of Construction~II}
In order to generalize Construction~II from Theorem~\ref{thm:construction_u_equal_q}, let us first generalize Theorem~\ref{thm:stuck-at-one-and-RRE}.
\begin{theorem}\label{thm:stuck-at-one-and-RRE_gen}
Let $q$ be a prime power and let a $\kappa \times n$ matrix $\H = \big(H_{i,j}\big)^{i \in [\kappa]}_{j\in [n]}$ over $\Fq$ be given.
Let $s = \max_{i \in [u]} s_i$.

If the RRE of any $\kappa\times u$ submatrix (denoted by $\H^{(u)}$) has the following form (up to column permutations:)
\begin{align}\label{eq:RRE_stuckatone_gen}
&\RRE(\H^{(u)}) = \\
&\begin{pmatrix}
\smash[b]{\blockRRE{ \leq \lceil q/s\rceil-1}} & \circ \;\circ \; \dots \; \circ  &\dots &\dots&\dots \; \circ\\
& \smash[b]{\blockRRE{ \leq \lceil q/s\rceil-1}} & \circ \;\circ \; \dots \; \circ& \dots&\dots\; \circ\\
&& \smash[b]{\blockRRE{ \leq \lceil q/s\rceil-1}}&\circ\; \dots& \dots\;\circ\\
&&& \ddots & \vdots\\
&\hspace{-10ex}\smash{\clap{\resizebox{0.35cm}{!}{$0$}}}&&& \smash[b]{\blockRRE{ \leq \lceil q/s\rceil-1}}\\
\end{pmatrix},\nonumber\\[-0.5ex]\nonumber
\end{align}
where $\bullet$ has to be a non-zero element from $\Fq$ and $\circ$ is any element from $\Fq$,
then, there exists a \PSMCTwo{u}{s} over $[q]$ of length $n$ and redundancy $r=\kappa$.
\end{theorem}
\begin{IEEEproof}
Follows from the proof of Theorem~\ref{thm:stuck-at-one-and-RRE_gen} with similar generalizations as in Theorem~\ref{thm:stuck-at-one-onerow_Di}.
\end{IEEEproof}
Notice that the largest $s_i$ restricts the size of the matrix and it is not clear how to consider the explicit values $s_i$ in this generalization and not only the maximum value.

Based on Theorem~\ref{thm:stuck-at-one-and-RRE_gen}, we can generalize Construction~II as follows.

\begin{theorem}[Generalized Construction~II]\label{thm:construction_u_equal_q_gen}
Given $\vec{s} = \vecelementsArb{s}{u} \in \intervallexcl{1}{q}^u$.
Let $s = \max_{i \in [u]} s_i$,
let $u \leq \lceil q/s\rceil+d-3$, $u \leq n$, $k < n$, and let $\H$ be a systematic $(n-k)\times n$ parity-check matrix of an $\codelinearHamming{n,k,d}$ code.
Then, there exists a \PSMCTwo{u}{\vec{s}} over $\Fq$ of length $n$ and redundancy $r=n-k$.
\end{theorem}
\begin{IEEEproof}
Similar to the proof of Theorem~\ref{thm:construction_u_equal_q} using Theorem~\ref{thm:stuck-at-one-and-RRE_gen}.
\end{IEEEproof}

\begin{corollary}\label{cor:generalized_constructionII}
If $q \geq 2$ is a prime power, then for all $u \leq n$ and $s \in \intervallexcl{1}{q}$,
there exists a
\PSMCTwo{u}{{s}} with redundancy $r=\max\{1,\redundECC{n}{u-\lceil q/s\rceil+3}{q}\}$ and therefore $\redundPSMC{n}{u}{s}{q} \leq \max \{1, \redundECC{n}{u-\lceil q/s\rceil+3}{q}\}$.
\end{corollary}

\subsection{Generalization of Construction~III}
In the following, we want to generalize Construction~III from Theorem~\ref{thm:constr_binary_code}.

In the first step, we suggest a generalization to
the case where all partially stuck-at cells are stuck at \emph{the same level} $s$, for some $1\leq s \leq q-1$.
For this purpose, we use a $Q$-ary code, where $Q \geq s+1$ is a prime power, and the principle consists of the following two steps:
\begin{enumerate}
\item Find $z \in [q]$ such that the number of the values contained in the set $\myset{S} = \{q-Q+1, \dots, q-1, 0, \dots, s-1\}$ of size $Q+s-1$ is  minimized in $({\vec{w}}^{(z)})_U = (\vec{w} + z \cdot \mathbf{1}_n)_U$.\\
Denote the number of values from $\myset{S}$ in $({\vec{w}}^{(z)})_U$ by $\widetilde{u}$.
\item Use a $Q$-ary \SMC{\widetilde{u}} to mask these $\widetilde{u}$ stuck cells.
\end{enumerate}
This leads to the following theorem.
\begin{theorem}[Generalized Construction~III for $s_i=s$]\label{thm:constr_binary_code_genD}
Let $n$, $q \geq 4$, $u$ be positive integers, let $s \in \intervallexcl{1}{q}$, and let $Q \geq s+1$ be a prime power.
Denote
\begin{equation*}
\widetilde{u} = \left\lfloor \frac{(Q+s-1)u}{q}\right\rfloor.
\end{equation*}
Assume that $\widetilde{\mycode{C}}$ is an $\codelinearHammingqInp{n,k,d\geq \widetilde{u}+1 }{Q}$ $Q$-ary \SMC{\widetilde{u}} with encoder $\widetilde{\mycode{E}}$ and decoder $\widetilde{\mycode{D}}$ given by Theorem~\ref{thm:usual-stuck-at}.
Then, there exists a \PSMCTwo{u}{s} over $[q]$ of length $(n+1)$ and redundancy
\begin{equation}\label{eq:red-gen3-sequal}
r =  (n-k-1)\log_q\left(\frac{q}{\lfloor q/Q\rfloor}\right) + 2.
\end{equation}
\end{theorem}
\begin{IEEEproof}
Let $\myset{S} = \{q-Q+1, \dots, q-1, 0, \dots, s-1\}$ of cardinality $Q+s-1$.
As in the proof of Theorem~\ref{thm:constr_binary_code}, we first show that there is some scalar $z$ such that $(\vec{w}^{(z)})_U$ has at most $\widetilde{u}$ entries from $\myset{S}$.
The value $\widetilde{u}$ is equivalent to the minimum number of any $Q+s-1$ consecutive values in $\vec{w}_U$. Denote $v_i = |\{j: w_j = i, j \in \{\phi_0, \dots, \phi_{u-1}\}|$, for all $i\in [q]$. Hence, we want to minimize $\sum_{j=0}^{Q+s-2} v_{i+j \!\mod\! q}$, $i \in [q]$.
Since $\sum_{i=0}^{q-1}\sum_{j=0}^{Q+s-2} v_{i+j \!\mod\! q} = (Q+s-1)u$, there exists an integer $i$ such that $\sum_{j=0}^{Q+s-2} v_{i+j \!\mod\! q} \leq \lfloor \frac{(Q+s-1)u}{q}\rfloor$ and therefore, there is a $z$ such that $({\vec{w}}^{(z)})_U$ contains at most $\widetilde{u} = \lfloor \frac{(Q+s-1)u}{q}\rfloor $ values from $\myset{S}$.

Second, we treat those $\widetilde{u}$ cells of $({\vec{w}}^{(z)})_U$, which have values in $\myset{S}$, like usual $Q$-ary stuck cells. Note that adding any value from the set $[Q]$ to the other $u-\widetilde{u}$ cells still masks the partially stuck-at-$s$ cells.
Therefore, we can use an $\codelinearHammingqInp{n,k,d\geq \widetilde{u}+1}{Q}$ code to mask these cells (see Theorem~\ref{thm:usual-stuck-at}).

As a generalization of Theorem~\ref{thm:constr_binary_code}, we choose $\vec{m} \in [q]^k$ and $\vec{m}^\prime \in \big[\left\lfloor q/Q\right\rfloor\big]^{n-k-1}$ and the statement on the redundancy follows.
\end{IEEEproof}
Note that if $Q+s-1 \geq q$, then $\widetilde{u} \geq u$ and instead of using Theorem~\ref{thm:constr_binary_code_genD}, we should use a $q$-ary \SMC{u} or another construction.

Let us now describe a method to generalize the construction from Theorem~\ref{thm:constr_binary_code} (respectively Theorem~\ref{thm:constr_binary_code_genD}) to the case where the cells are partially stuck-at different levels, given by $\vec{s} = \vecelementsArb{s}{u}$.
We want to use a $Q$-ary SMC, where $Q \geq \max_i\{s_i\}+1$, to obtain an \PSMCTwo{u}{\vec{s}}. Of course, one way is to define $s = \max_{i} \{s_i\}$ and to consider the cells as partially stuck-at-$s$ cells and use Theorem~\ref{thm:constr_binary_code_genD}.
However, we can do better by refining the "minimization step" as done in the following theorem.

\begin{theorem}[Generalized Construction~III]\label{thm:constr_binary_code_gen_si}
Let $n$, $q \geq 4$ and $u$ be positive integers.
Let $\vec{s} = \vecelementsArb{s}{u} \in \intervallexcl{1}{q}^u$ be given and denote $u_i = |\{s_j = i, \forall j \in [u]\}|$, $\forall i \in [q]$. Then, $u = \sum_{i=1}^{q-1} u_i$ and let $Q \geq \max_i\{s_i\}+1$ be a prime power.
Denote
$\sigma_i= \min\{q,Q+i-1\}$, $\forall i \in [q]$, and
\begin{equation*}
\widetilde{u} = \left\lfloor \frac{\sum_{i=1}^{q-1 }u_{i} \cdot \sigma_{i}}{q}\right\rfloor.
\end{equation*}
Assume that $\widetilde{\mycode{C}}$ is an $\codelinearHammingqInp{n,k,d\geq \widetilde{u}+1 }{Q}$ $Q$-ary \SMC{\widetilde{u}} with encoder $\widetilde{\mycode{E}}$ and decoder $\widetilde{\mycode{D}}$ given by Theorem~\ref{thm:usual-stuck-at}.
Then, there exists a \PSMCTwo{u}{\vec{s}} over $[q]$ of length $(n+1)$ and redundancy
\begin{equation}\label{eq:red-gen3-sdiff}
r =  (n-k-1)\log_q\left(\frac{q}{\lfloor q/Q\rfloor}\right) + 2.
\end{equation}
\end{theorem}
\begin{IEEEproof}
If a cell in $\myset{U}_{i}$ has a cell level in $\intervallexcl{1}{q}$, then this partially stuck-at cell is still masked after adding any value from $[Q]$.
Thus, we want to find $z \in [q]$ such that the partially stuck-at positions in the vector $({\vec{w}}^{(z)})_{U_i}=(\vec{w} + z \cdot \mathbf{1}_{n+1})_{\myset{U}_{i}}$ contain as many values in $\intervallincl{i}{q-Q}$, $\forall i \in \intervallexcl{1}{q}$, as possible.
Equivalently, we want to minimize the number of values from $[q]\setminus \intervallincl{i}{q-Q}$ in $({\vec{w}}^{(z)})_{U_i}$, for all $i \in \intervallexcl{1}{q}$.

In the sequel, we describe a way how accomplish this. We have
$$\sigma_i \defeq \big|[q]\setminus \intervallincl{i}{q-Q}\big| = \min\{q,Q+i-1\}, \ \forall i \in [q],$$
and generalizing Theorems~\ref{thm:constr_binary_code} and~\ref{thm:constr_binary_code_genD}, we define:
\begin{equation}\label{eq:definition_vellli}
v_\ell^{(i)} = |\{j: w_j = \ell, j \in \myset{U}_{i}\}|, \ \forall \ell\in [q],  \ i \in \intervallexcl{1}{q}.
\end{equation}
We want to minimize (subject to $\ell \in [q]$)
\begin{equation*}
\sum_{i=1}^{q-1 }\sum\limits_{j=0}^{\sigma_i-1} v^{(i)}_{\ell+j \mod q}.
\end{equation*}
We know that
\begin{equation*}
\sum_{\ell=0}^{q-1 }\sum_{i=1}^{q-1 }\sum\limits_{j=0}^{\sigma_i-1} v^{(i)}_{\ell+j \mod q} = \sum_{i=1}^{q-1 }u_{i} \cdot \sigma_{i}.
\end{equation*}
Thus, by the pigeonhole principle, there exists an integer $\ell \in [q]$ such that
\begin{equation*}
\sum_{i=1}^{q-1 }\sum\limits_{j=0}^{\sigma_i-1} v^{(i)}_{\ell+j \mod q} \leq \left\lfloor \frac{\sum_{i=1}^{q-1 }u_{i} \cdot \sigma_{i}}{q}\right\rfloor \defeq \widetilde{u}.
\end{equation*}
Similar to Theorems~\ref{thm:constr_binary_code} and~\ref{thm:constr_binary_code_genD}, we can treat these $\widetilde{u}$ partially stuck-at cells as $Q$-ary cells which are usually stuck-at.
With a similar proof as in Theorems~\ref{thm:constr_binary_code} and~\ref{thm:constr_binary_code_genD}, this leads to the statement.
\end{IEEEproof}
\begin{example}

Let $n=31$, $q = 8$, and $\vec{s}=(1 \ 1 \ 1 \ 1 \ 2 \ 2 \ 3)$. Hence, $u_1=4$, $u_2=2$, $u_3=1$  and $u = 7$. We choose $Q = 4$ and we want to use a $4$-ary SMC to construct a \PSMCTwo{u}{\vec{s}} of length $32$.

The first approach is to consider all cells as partially stuck-at level $s = \max_i \{s_i\} = 3$. Then, with Theorem~\ref{thm:constr_binary_code_genD} and $Q = 4$, we have $\widetilde{u} =  5$ and we need a code over $\F_{2^2}$ of length $31$ and distance at least $6$. The largest known such code is a $\codelinearHammingqInp{31,23,6}{4}$ code (see \cite{Grassl:codetables}) and therefore, with~\eqref{eq:red-gen3-sequal}, the required redundancy is $6.66$.

Let us now compare this to Theorem~\ref{thm:constr_binary_code_gen_si}.
We denote $\sigma_1 = 4$, $\sigma_2 = 5$ and $\sigma_3 = 6$ and therefore,
\begin{equation*}
\widetilde{u} =  \left\lfloor \frac{4u_1 + 5 u_2 + 6 u_3}{q}\right\rfloor = 4.
\end{equation*}
We need a code over $\F_{2^2}$ of length $31$ and distance at least $\widetilde{u}+1=5$. The largest such known code is a $\codelinearHammingqInp{31,24,5}{4}$ code (see~\cite{Grassl:codetables}) and with~\eqref{eq:red-gen3-sdiff}, the required redundancy is $r =6$, i.e., compared to the approach of Theorem~\ref{thm:constr_binary_code_genD}, we have decreased the redundancy.

The generalized Construction~II from Theorem~\ref{thm:construction_u_equal_q_gen} needs redundancy $\redundECC{32}{7}{8} = 10$.
Further, the upper bound from Theorem~\ref{thm:bounds_stuckatsi}, gives $7.01$ and we also improve upon this bound.

\end{example}

\section{Codes for Cells with Unreachable Levels}\label{sec:unreachable}
In flash memories, it might happen that certain levels cannot be reached or should not be programmed anymore since they are highly unreliable, see e.g.~\cite{Gabrys2014Coding}. Finding codes that mask these cells can be seen as the dual problem to finding PSMCs.
Namely, we want to find a code as follows.
\begin{definition}[Codes for Unreachable Levels]
An $(n,M)_q$ \textbf{${(u, \vec{s})}$-unreachable-masking code} (\UnreachTwo{u}{\vec{s}})~$\mycode{C}$  is a coding scheme with encoder $\mathcal{E}$ and decoder $\mathcal{D}$. The input to the encoder~$\mathcal{E}$ is the set of locations $\{\phi_0, \dots, \phi_{u-1}\}\subseteq [n]$, the unreachable levels $\vec{s} = \vecelementsArb{s}{u}\in [q-1]^{u}$ of some $u\leq n$ cells and a message $m\in [M]$. Its output is a vector $\vec{y}^{(m)}\in [q]^n$ which masks the values of the~$u$ cells with unreachable levels, i.e., $$y_i^{(m)} \leq s_i, \forall i \in \{\phi_0,\phi_1,\dots,\phi_{u-1}\},$$
and its decoded value is $m$, that is $\mathcal{D}(\vec{y}^{(m)}) =m$.
\end{definition}

The following theorem establishes a connection between PSMCs and UMCs.
\begin{theorem}~\label{lem:dualcode}
Given a \PSMCTwo{u}{\vec{s}} with $\vec{s} = (q-1-s_0 , q-1-s_1 , \dots , q-1-s_{u-1})$ and redundancy $r$.
Then, this code can be used as a \UnreachTwo{u}{\widetilde{\vec{s}}} with  $\widetilde{\vec{s}} = (s_0 , s_1 , \dots , s_{u-1})$ and redundancy $r$.
\end{theorem}
\begin{IEEEproof}
Assume we want to write a message into $n$ memory cells where the levels at positions $\phi_0,\phi_1,\dots,\phi_{u-1}$ can be at most $s_0,s_1,\dots,s_{u-1}$. We construct a \PSMCTwo{u}{\vec{s}} with redundancy $r$ for the levels $\vec{s} = (q-1-s_0 , q-1-s_1 , \dots , q-1-s_{u-1})$. The output of this encoder is a vector $\vec{y}^{(m)}$ such that $y^{(m)}_i \geq q-1-s_i$, $\forall i \in \{\phi_0,\phi_1, \dots, \phi_{u-1}\}$.

Therefore, the vector $\bar{\vec{y}^{(m)}} \defeq \vecelements{q-1-y^{(m)}}$ has the property that
$$\bar{y^{(m)}_i}  \leq q-1-(q-1-s_i)=s_i,\quad \forall i \in \{\phi_0, \dots, \phi_{u-1}\},$$
and therefore, we write $\bar{\vec{y}^{(m)}}$ into the cells and have constructed a \UnreachTwo{u}{\widetilde{\vec{s}}} with  $\widetilde{\vec{s}} = (s_0 , s_1 , \dots , s_{u-1})$ and redundancy $r$.
\end{IEEEproof}

The construction by Gabrys, Sala and Dolecek from~\cite{Gabrys2014Coding} is based on tensor product codes and can mask unreliable cells and correct additional random errors.
To compare their construction to our results, we assume that no random errors occur (i.e., in their notation $t_1=t_2=0$).
Then, their matrix $H_3$ defines an $[n,n,0]_4$ code (in their notation, meaning it does not correct any error), $H_4$ defines an $[n,k_4,0,\ell]_2$ code (meaning it is a \SMC{\ell} which cannot correct additional random errors). Therefore, $d_4 \geq \ell+1$, see e.g., \cite{Heegard-PartitionedLinearBlockCodesStuckAtDefects_1983} and due to the Singleton bound $k_4 \leq n-d_4+1 = n-\ell$.
The construction from~\cite[p.~1492]{Gabrys2014Coding} yields a code for masking the unreliable cells of length $3n$ and dimension $2k_3+k_4 \leq 3n- \ell$. Thus, their required redundancy is $r \geq \ell$, which is the same as the one for a \SMC{\ell} and thus, their construction does not give any improvement compared to \cite{Heegard-PartitionedLinearBlockCodesStuckAtDefects_1983}.
As shown in the previous sections and in Theorem~\ref{lem:dualcode}, our constructions therefore improve on both, \cite{Heegard-PartitionedLinearBlockCodesStuckAtDefects_1983} and \cite{Gabrys2014Coding}, when masking memory cells with unreachable levels.

\section{The Capacity of the Partially Stuck-at Cell Channel}\label{sec:capacity}
In this section, we study the setup of partially stuck-at cells from the capacity point of view. For fixed integers $q>1$ and $0<s<q$, we consider the storage channel in which a cell can be partially stuck-at level $s$ with probability $p$. This channel model is an example of a discrete memoryless memory cell which was studied by Heegard and El Gamal in~\cite{Heegard-El-Gamal_1983}. In the partially stuck-at cell model, we assume that a cell is partially stuck-at level $s$ with some probability $p$. If a letter $x\in X = \{0,1,\ldots,q-1\}$ is stored in a cell, then the letter $y\in Y = \{0,1,\ldots,q-1\}$ is retrieved where $y = \max \{x,s\}$ with probability $p$.

If \emph{both}, the encoder and decoder, know the state of the memory, i.e. the locations and levels of the partially stuck-at cells, then the capacity of this channel is
\begin{equation}\label{eq:capacity}
C_q(p,s) = 1-p+p\log_q(q-s) = 1-p\log_q\left( \frac{q}{q-s}  \right).
\end{equation}
To see this, notice that the lower bound on the redundancy from Theorem~\ref{thm:bounds_stuckatsi} holds also in this case as this is the number of all different memory states that can be programmed.
The capacity is achievable since both the encoder and decoder know the locations of the partially stuck-at memory cells and thus can use all programmable memory states.
Furthermore, according to~\cite{Heegard-El-Gamal_1983}, the capacity of any discrete memoryless memory cell channel in the case where \emph{only the encoder} knows the memory state is the same as the one where both the encoder and decoder know the memory state. Thus, we conclude that~(\ref{eq:capacity}) is the capacity of the model studied in this work.

In the following, we want to analyze how close our constructions are to the capacity.
Let us first consider the case $s=1$. The construction based on binary codes from Section~\ref{sec:binary construction} can mask $u=pn$ partially stuck-at-$1$ memory cells if there exists an \codelinearHammingqInp{n,k,d}{2} binary code that can mask $\widetilde{u} = \lfloor{2u}/{q}\rfloor = \lfloor{2pn}/{q}\rfloor$ binary usual stuck-at cells (which is the case if $d \geq \widetilde{u}+1$, see Theorem~\ref{thm:usual-stuck-at}).
Assume there exists a family of capacity-achieving linear binary codes with these properties, then we can use this family of codes in order to construct a family of PSMCs.
Asymptotically, for large $n$ and any $q$, the maximum achievable code rate of the construction from Theorem~\ref{thm:constr_binary_code} approaches the value
\begin{align}
R_q(p,1) &= \frac{n-r}{n} = 1 - \frac{2p}{q}\log_q\left(\frac{q}{\lfloor q/2 \rfloor}\right).\nonumber
\end{align}

Similarly, for arbitrary $s$ (for simplicity assume that $Q = s+1$ is a power of a prime), we obtain a family of PSMCs whose maximum achievable rate approaches the value
$$R_q(p,s) = 1-\frac{2sp}{q}\log_q\left(\frac{q}{\lfloor q/(s+1)\rfloor}\right).$$

Finally, we conclude that the difference between the capacity and the redundancy of this construction is given by
\begin{align}
 C_q&(p,s) - R_q(p,s)\nonumber \\
& =p\cdot \underbrace{\left( \frac{2s}{q}\log_q\left(\frac{q}{\lfloor q/(s+1)\rfloor}\right) - \log_q\left( \frac{q}{q-s}\right) \right)}_{\defeq\Delta(q,s)}, \label{eq:difference_capred}
\end{align}
where we call $\Delta(q,s)$ the \emph{difference coefficient}.

Table~\ref{tab:difference_coeffcient_capacity} shows the values of the difference coefficient $\Delta(q,s)$ for different values of $q$ and $s$.
\begin{table}[htb]
\centering
\caption{The difference coefficient $\Delta(q,s)$ for different values of $q $ and $s$, where $0 <s <q$ where $s+1$ is a prime power.}
\label{tab:difference_coeffcient_capacity}
\begin{tabular}{c|cccccc}
&$s=1$&$s=2$&$s=3$&$s=4$& $s=6$&$s=7$\\
\hline\\[-1.5ex]
$q = 2$ & 0	&	&	&	&\\
$q=3$ & 0.29	&0.33\\
$q=4$ & 0.042&	0.5&	0.5\\
$q=5$ & 0.089& 	0.48 &	0.63&	0.6\\
$q=6$ & 0.027 & 0.18 &	0.61&	0.72\\
$q=7$ & 0.045 &	0.19 & 0.57 &	0.71 &	0.71 \\	
$q=8$ & 0.019 &	0.19 &	0.27 &	0.67 &	0.83 &	0.75\\
$q=11$ & 0.020 &	0.11&	0.25&	0.33 &	0.76& 0.85\\
$q=13$ & 0.015&	0.076	&0.16	& 0.31 &	0.68 &	0.77\\ 
$q=16$&	0.0079&	0.057&	0.11&	0.19&	0.39&	0.45\\ 
$q=21$&	0.0072&	0.036&	0.084&	0.14&	0.25&	0.38\\ 
$q=32$&	0.0033&	0.023&	0.047&	0.082&	0.17&	0.19\\ 
\end{tabular}
\end{table}

\begin{figure}[htb]
\centering
\hspace{-4ex}
\includegraphics{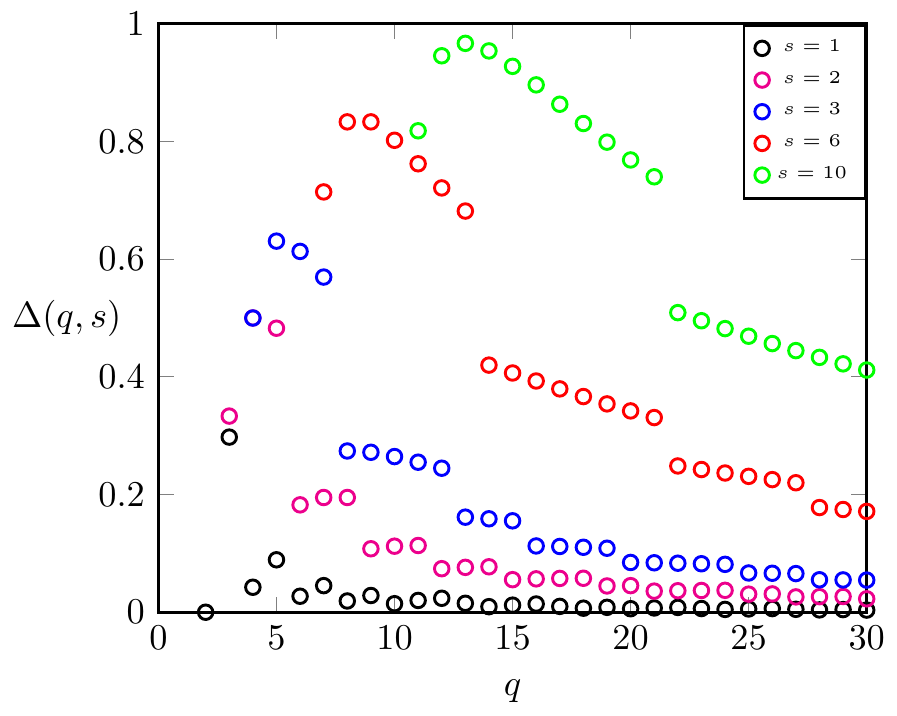}
%
\caption{The difference coefficient $\Delta(q,s)$ for different values of $q$ and $s$.}
\label{fig:difference_coefficient}
\end{figure}

In the same way, Figure~\ref{fig:difference_coefficient} illustrates the difference coefficient for some values of $q$ and~$s$.
We see that the difference coefficient tends to zero for $q$ large enough. The "plateaus" in the plots occur due to the floor operation in~\eqref{eq:difference_capred}.
A similar analysis can be done for the \PSMCTwo{u}{\vec{s}} construction from Theorem~\ref{thm:constr_binary_code_gen_si}.

For the special case $s=1$ and when $q$ is large enough, we neglect the floor operation and we can use the approximation $\ln(q)- \ln(q-1) \approx 1/q$ and obtain
\begin{align}
C_q(p,1)&-R_q(p,1) =\nonumber\\
& \approx p\left( \frac{2\log_q(2)}{q} - \log_q\left( \frac{q}{q-1}\right) \right)\nonumber\\
& \approx p\left( \frac{2\ln(2)}{q \ln(q)} - \frac{1}{q \ln(q)} \right)
\approx 0.38\frac{p}{q\ln(q)}.\label{eq:diff-capa-rate-s=1}
\end{align}
Thus, we can evaluate how close the code rate from Theorems~\ref{thm:constr_binary_code} and \ref{thm:constr_binary_code_genD} come to the capacity from~\eqref{eq:capacity}.
Further,~\eqref{eq:diff-capa-rate-s=1} shows that asymptotically, the coefficient $\Delta(q,1)$ approaches zero when $q$ increases. In the same way, for fixed $s$, the coefficient $\Delta(q,s)$ approaches 0 with increasing $q$.

We note that the trivial construction from Theorem~\ref{thm:bounds_stuckatsi}, in which only the levels $s,\ldots, q-1$ are used, can also be used to mask any number of cells which are partially stuck-at-$s$. The rate of this construction is $\log_q(q-s)$, and it is greater than the value of $R_q(p,s)$, when $q$ is a multiple of $s+1$, if
$$\log_q(q-s) \geq 1-\frac{2sp}{q}\log_q(s+1),$$
or
$$p\geq \frac{q}{2s}\log_{s+1}\left( \frac{q}{q-s} \right).$$
For example, for $s=1$ we get that this threshold equals
$$\frac{q}{2}\log\left( \frac{q}{q-1} \right)$$
 and for $q$ large enough, this value approaches $1/(2\ln 2) \approx 0.7213$.

\begin{figure}[htb]
\centering
\hspace{-2.5ex}
\includegraphics{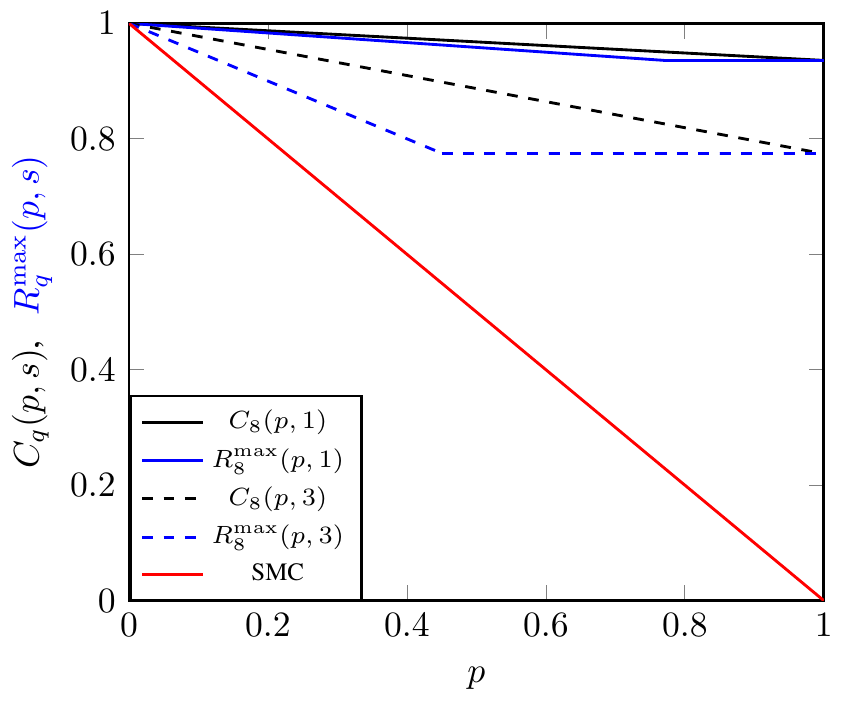}
%
\caption{Capacity $C_q(p,s)$ and maximum achievable rate $R_q^{\max}(p,s)$ for $q=8$.}
\label{fig:C-R_q8}
\end{figure}

\begin{figure}[htb]
\centering
\hspace{-4ex}
\includegraphics{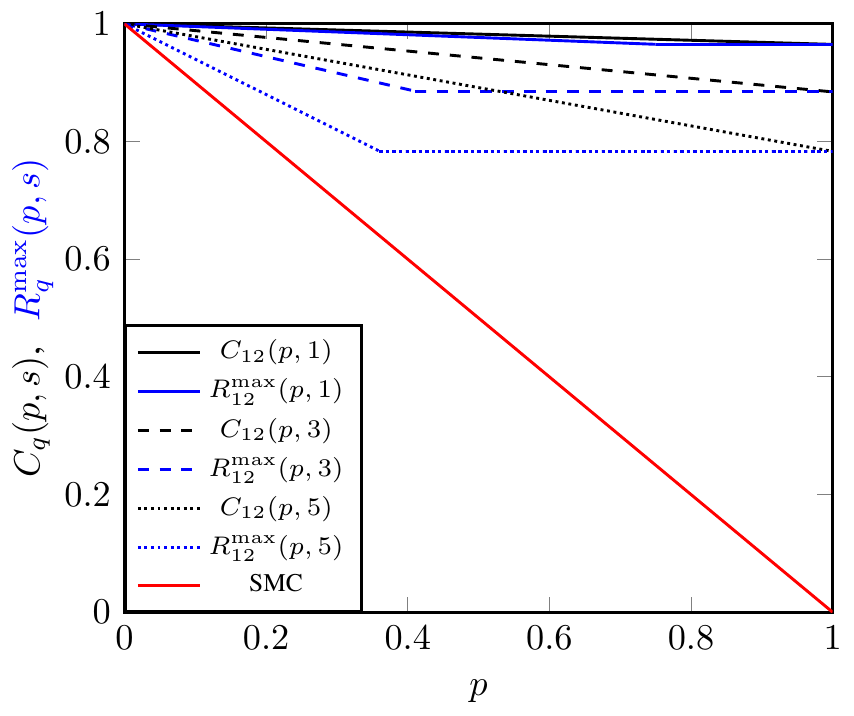}
\caption{Capacity $C_q(p,s)$ and maximum achievable rate $R^{\max}_q(p,s)$ for $q=12$.}
\label{fig:C-R_q12}
\end{figure}

We conclude that the maximum achievable rates, denoted by $R_q^{\max}(p,s)$, according to the constructions from Theorems~\ref{thm:bounds_stuckatsi},~\ref{thm:constr_binary_code} and~\ref{thm:constr_binary_code_genD}, when $q$ is a multiple of $s+1$, is given by
\begin{displaymath}
R_q^{\max}(p,s)= \left\{ \begin{array}{ll}
1-\frac{2sp}{q}\log_q(s+1) & \textrm{if $p\leq \frac{q}{2s}\log_{s+1}\left( \frac{q}{q-s} \right)$}\\
\log_q(q-s) & \textrm{else}
\end{array} \right.
\end{displaymath}
In Figures~\ref{fig:C-R_q8} and~\ref{fig:C-R_q12}, we plot the graphs of the capacity $C(p,s)$ and the maximum achievable rates $R_q^{\max}(p,s)$ for $q=8, s=1,3$ and $q=12, s=1,3,5$.

Notice that asymptotically, both an SMC and Construction~II have rate $R = 1-p$ and are therefore worse than $R_q^{\max}(p,s)$.

\section{Conclusion}\label{sec:conclusion}
In this paper, we have considered codes for masking partially stuck-at memory cells. After defining the defect model, we have derived lower and upper bounds on the minimum redundancy which is required to mask partially stuck-at cells in multilevel memories.
We have presented three constructions of \emph{partially stuck-at masking codes} (PSMCs). The first one masks any $u<q$ partially stuck-at cells, where $q$ is the number of cell levels, and requires less than one redundancy symbol. When the cells are partially stuck-at level $1$ (i.e., $s=1$), this construction is asymptotically optimal. The second construction is based on the parity-check matrix of an error-correcting code and works well for $u \geq q$ (where $q$ is a prime power).
The last construction is based on using a binary error-correcting code and can be applied for any $u$ and $q$.
The three constructions were first derived for $s=1$ and then generalized to arbitrary stuck levels.
We have further shown how these PSMCs can be applied for cells with unreachable levels (which can be seen as the dual problem to partially stuck-at cells) and that they outperform known code constructions for this model.
Further, we have considered the capacity of the partially stuck-at channel and have analyzed the gap between the capacity and the achieved code rate. For $s=1$, we achieve the capacity asymptotically.

The following table for $s=1$ provides an overview of our constructions and their required redundancies.
Recall that $\redundECC{n}{d}{q}$ denotes the smallest redundancy of a linear error-correcting code of length $n$ and minimum Hamming distance~$d$ over $\Fq$.

\begin{center}
\begin{tabular}{p{0.06\textwidth}|p{0.37\textwidth}}
 & Upper bound on $\redundPSMC{n}{u}{\mathbf{1}}{q}$\\[0.2ex]
\hline\hline\\[-1ex]
$u <q$ & $1-\log_q\left\lfloor \frac{q}{u+1} \right\rfloor$ (Theorem~\ref{thm:stuck-at-one-onerow})\\[2ex] \hline\\[-1ex]
$u \geq q$ & $\redundECC{n}{u-q+3}{q}$ (Corollary~\ref{cor:constructionII})\\[2ex] \hline\\[-1ex]
any~$u,q$ & $\left(\redundECC{n-1}{\lfloor \frac{2u}{q}\rfloor+1}{2}-1\right)\cdot\log_q\left(\frac{q}{\lfloor q/2\rfloor}\right) + 2$ \\[2ex]
& (Theorem~\ref{thm:constr_binary_code})
\end{tabular}
\end{center}
\vspace{1ex}
Similarly, the following table gives an overview of the constructions for arbitrary partially stuck-at levels, given by a vector $\vec{s} = \vecelementsArb{s}{u} \in \intervallexcl{1}{q}^u$.\\
\begin{center}
\begin{tabular}{p{0.12\textwidth}|p{0.32\textwidth}}
 & Upper bound on $\redundPSMC{n}{u}{\mathbf{s}}{q}$\\[0.2ex]
\hline\hline\\[-1ex]
$\sum\limits_{i=0}^{u-1}s_i < q$ & $r = 1-\log_q\left\lfloor\frac{q}{\sum_{i=0}^{u-1}s_i+1}\right\rfloor$ (Theorem~\ref{thm:stuck-at-one-onerow_Di})\\[2ex] \hline\\[-1ex]
$u \geq \big\lceil\frac{q}{\max_i s_i}\big\rceil$ & $\redundECC{n}{u-\Big\lceil \frac{q}{\max_i s_i}\Big\rceil+3}{q}$ (Corollary~\ref{cor:generalized_constructionII})\\[2ex] \hline\\[-1ex]
any~$u,q$ & $\left(\redundECC{n-1}{\left\lfloor \frac{\sum_{i=1}^{q-1 }u_{i} \cdot \sigma_{i}}{2}\right\rfloor+1}{Q}-1\right)$\newline\phantom{AAAAAAAAAAAA}$\cdot\log_q\left(\frac{q}{\lfloor q/Q\rfloor}\right) + 2$, \\[2ex]
& where $Q \geq \max_i\{s_i\}+1$ is a prime power and
$\sigma_i= \min\{q,Q+i-1\}$, $\forall i \in [q]$. (Theorem~\ref{thm:constr_binary_code_gen_si})
\end{tabular}
\end{center}
\vspace{1ex}

Thus, we have provided codes for masking partially stuck-at cells for any set of parameters and have shown that the required redundancy significantly decreases compared to SMCs and to a trivial construction.
\section*{Acknowledgement}
A.~Wachter-Zeh's work has been supported in part by a Minerva Postdoctoral Fellowship (Max-Planck Society) and in part by the Marie Sk\l{}odowska-Curie Individual Fellowship "DARE" (grant number 655109) from the European Commission within the Horizon 2020 program. E. Yaakobi's work has been supported in part by the Israel Science Foundation (ISF) grant No. 1624/14.

\bibliographystyle{IEEEtranS}
\bibliography{antoniawachter}
\end{document}